\documentclass[9pt,twocolumn]{extarticle}

\usepackage[a4paper, left=0.75in, right=0.75in, top=0.75in, bottom=0.75in]{geometry}
\usepackage[dvipsnames]{xcolor}      
\usepackage{listings}    
\usepackage{setspace}    
\usepackage{titlesec}    
\usepackage[colorlinks=true, linkcolor=black, urlcolor=gray, citecolor=gray]{hyperref} 
\usepackage[numbers,sort&compress]{natbib} 
\usepackage{cuted}
\usepackage{graphicx}     
\usepackage{caption}
\usepackage{mathptmx}   
\usepackage{braket}
\usepackage{authblk}  
\usepackage{dblfloatfix}  
\usepackage{bm}
\usepackage{stmaryrd}
\usepackage{amsmath}
\usepackage{newpxtext}
\usepackage[colorlinks=true, linkcolor=black, urlcolor=gray, citecolor=gray]{hyperref} %
\usepackage[normalem]{ulem}
\usepackage{xcolor}
\usepackage{graphicx}
\usepackage{sidecap}
\usepackage{bibunits}
\usepackage{booktabs}
\usepackage{multirow}
\usepackage{makecell}
\usepackage{amssymb,stackengine,graphicx}

\renewenvironment{abstract}
{%
  \begin{strip}
  \par\noindent
  {\color{blue!70!black}\rule{\textwidth}{3pt}\par}\large \rmfamily
  \vspace{2.0em}
  \hfill
  \begin{minipage}{0.8\textwidth}
}
{%
  \end{minipage}\par
  \vspace{2.0em}
  {
    }
  \end{strip}
}

\titleformat{\section}
  {\color{black}\normalfont\large\bfseries}
  {}
  {0em}
  {} 

  \titleformat{\subsection}
  {\color{black}\normalfont\Large\bfseries} 
  {}
  {0em}
  {} 

\newif\ifshowcomments\showcommentstrue

\newcommand{\cdag}{\hat{c}^{\dagger}}
\renewcommand{\c}{\hat{c}^{\phantom{\dagger}}}

\newcommand{\bairo}{Ba$_2$IrO$_4$}
\newcommand{\sriro}{Sr$_2$IrO$_4$}

\newcommand{\jeff}{$j_{\mathrm{eff}}=1/2$}
\newcommand{\JEFF}{$j_{\mathrm{eff}}=3/2$}

\makeatletter
\newcommand{\fixed@sra}{$\vrule height 2\fontdimen22\textfont2 width 0pt\shortrightarrow$}
\newcommand{\shortarrow}[1]{%
  \mathrel{\text{\rotatebox[origin=c]{\numexpr#1*45}{\fixed@sra}}}
}
\makeatother

\title{{\Huge \textbf{Where Slater meets Mott: the insulating phase of the 214 Ruddlesden-Popper iridates}}}

\author[1,2]{Francesco Cassol}
\author[2]{Michele Casula}
\author[2]{Benjamin Lenz}

\affil[1]{Centre de Physique Théorique, CNRS, Ecole Polytechnique, 
Institut Polytechnique de Paris, F-91128 Palaiseau, France}
\affil[2]{Institut de Min\'eralogie, de
  Physique des Mat\'eriaux et de Cosmochimie, Sorbonne Universit\'e, CNRS UMR7590, MNHN, 4 Place Jussieu, 75005, Paris, France}

\date{\today}

\begin{document}

\maketitle

\begin{abstract}
The $5d^5$ Ruddlesden-Popper iridates \sriro\ and \bairo\ have attracted
considerable attention as spin-orbit entangled analogues of the cuprates,
yet the origin of their antiferromagnetic insulating state, of Slater
or of Mott-Heisenberg type, remains debated.
We find that both materials sit in an intermediate regime in which itinerant
and localized electrons cooperate in stabilizing the antiferromagnetic order,
with \bairo\ slightly more correlated than \sriro.
We obtain this by computing, within dynamical mean-field theory, the
interaction-temperature phase diagram of a realistic three-orbital $t_{2g}$
Hubbard model with spin-orbit coupling.
Our study shows that neither the behavior of the gap across the N\'eel temperature nor the
spin-polaron features of the spectral function unambiguously identify the
cooperative regime, whereas the two-peak structure of
the optical conductivity is reproduced only there and therefore constitutes a reliable fingerprint.
By placing both compounds in a common phase diagram, our results reconcile
seemingly contradictory experimental and theoretical reports, and provide a
unified framework for future studies of the 214 iridates and of other
spin-orbit coupled correlated materials.
\end{abstract}


Antiferromagnetic insulators are traditionally understood in terms of two
limiting pictures. In the Slater scenario~\cite{Slater_1951}, valid at weak
electronic interaction, the insulating state is a consequence of
magnetic order.
The gap is proportional to the staggered magnetization, the state above the N\'eel
temperature $T_N$ is a paramagnetic metal without well-formed local moments,
and $T_N$ increases with the local interaction strength $U$.
In the opposite
Mott-Heisenberg limit, the charge gap is of order $U$ and exists independently
of magnetic order. Local moments are already formed at temperatures far above
$T_N$, and ordering merely reflects the freezing of their orientation under
the superexchange coupling $J \sim 4t^2/U$, with $t$ the effective
hopping parameter, so that $T_N$ decreases with
increasing $U$. The low-energy physics is then captured by a Heisenberg model,
and the paramagnetic phase above $T_N$ is a Mott insulator with fluctuating
moments.

Theory-wise, the two regimes are not distinct phases and thorough discussions are often based on the
half-filled single-band Hubbard model on a bipartite lattice.
There, the ordered states share the same broken symmetry and are connected by a smooth crossover,
signalled by the non-monotonic $T_N(U)$ found for instance in dynamical mean-field theory (DMFT) and its extensions~\cite{Pruschke_2003,Sangiovanni_2006}.
Quantitative criteria to discriminate the two regimes have therefore been sought.
One relies upon the energetic origin of the ordering,
which is driven by a gain in potential energy on the Slater side and in
kinetic energy on the Heisenberg
side~\cite{Taranto_2012,Fratino_2017}.
Another is based on the nature of the single-particle and spin excitations, with coherent
Bogoliubov-like quasiparticles and Landau-damped magnons on one side~\cite{Chubukov_1992}, and
Hubbard bands with spin-polaron features on the other~\cite{Sangiovanni_2006}.
These criteria do not need to coincide, and the
intermediate regime where most materials of interest lie, i.e. $U \sim
W$, with $W$ the bandwidth, remains poorly characterized.

In multi-orbital systems this
dichotomy is further complicated by the interplay of Hund's coupling, orbital degeneracy, crystal-field splitting and spin-orbit coupling (SOC).
The 214 Ruddlesden-Popper iridates illustrate the difficulty.
They feature an Ir $5d^{5}$ electronic configuration, where
the SOC lifts the degeneracy of the
$t_{2g}$ manifold into $j_{\mathrm{eff}}=1/2$ and
$j_{\mathrm{eff}}=3/2$ states.
While the materials' small gap insulating phase can be explained by a
metal-insulator transition which mainly involves the half-filled
\jeff\ band~\cite{Kim_2008,Martins_2011,Arita_2012,Pavarini_2023,Cassol_2024},
there is a lack of consensus on the nature of the lower-temperature
magnetic phase. Some theoretical
studies support a Slater-type antiferromagnet~\cite{Arita_2012,Li_2013,Seki_2013, Choi_2024}, others favor a Mott-like
picture~\cite{Martins_2011,Moser_2014,Pavarini_2021,Foulquier_2023,Pavarini_2025},
and yet others explicitly point to a mixed behavior~\cite{Watanabe_2014}.

\begin{figure*}[t!]
    \centering
    \begin{minipage}{0.62\linewidth}
        \centering
        \includegraphics[width=\linewidth]{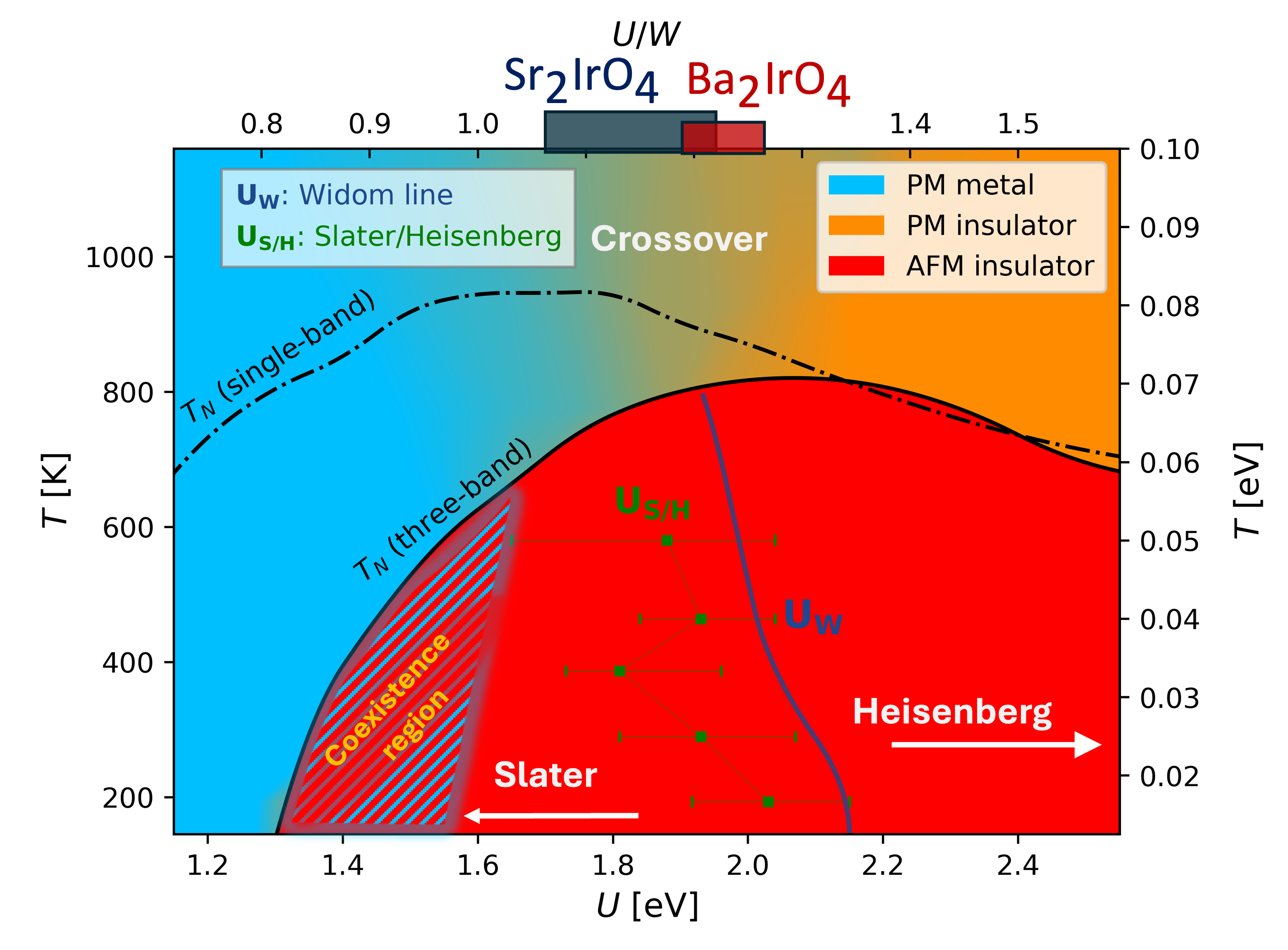}
    \end{minipage}
    \hfill
    \begin{minipage}{0.30\linewidth}
    \caption{$U-T$ phase diagram of $t_{2g}^5$ iridates.
    The solid (dashed-dotted) black line represents the transition from a paramagnetic to a magnetic phase in the three(single)-band model.
    The blue line $U_W$ denotes the Widom line within the fully paramagnetic solution. 
    The green line marks the passage from Slater to Heisenberg magnetism. The horizontal green bars denote a region where magnetism is stabilized by the cooperation of localized and itinerant electrons.
    The color scheme refers to phases in the three-band model. Within the AFM dome, the shaded area
    delimits the coexistence region. 
    The upper axis reports the Coulomb interaction in units of the bandwidth of the \jeff\ band $W$.
    Blue and red bars show the realistic interaction ranges for Sr$_2$IrO$_4$ and Ba$_2$IrO$_4$, respectively.
     }
    \label{fig:Figure_1}
    \end{minipage}
        \vspace{5pt}
    \hrule
\end{figure*}

Experimentally, the distinction relies on a combination of probes, none of
which is conclusive on its own. These include transport and optical
conductivity across $T_N$ to detect the presence or absence of a metal-insulator transition
concomitant with magnetic order~\cite{Moon_2009,Seo_2017}; photoemission and tunnelling
spectroscopy to check the persistence of the gap above
$T_N$~\cite{Kim_2008,Li_2013,Moser_2014,Dai_2014,Uchida_2014}; magnetic susceptibility
and neutron or resonant x-ray scattering to study the magnetic
behavior~\cite{Kim_2009,JKim_2012,Fujiyama_2012}; and time-resolved spectroscopies that
separate electronic from magnetic time scales~\cite{Hsieh_2012}.
Despite deploying the full experimental toolkit, both scenarios have
been supported~\cite{Hsieh_2012}: although no gap closure occurs above
the Néel temperature $T_N$ of approximately 250 K~\cite{Moser_2014,
  Uchida_2014}, supporting a Mott picture, a strong temperature
dependence is observed in the optical conductivity~\cite{Moon_2009,Seo_2017}, along with a gap softening when approaching
$T_N$~\cite{Uchida_2014}, 
suggestive of a Slater insulator, whose charge gap depends on the spin density wave order.
Even though it is experimentally established that the $j_{\mathrm{eff}}=3/2$ states are completely filled~\cite{Moser_2014,Uchida_2014,DelaTorre_2015}, it is less clear to which extent they can be ignored when describing the low-energy physics. 
For instance, recent literature has emphasized that a multi-orbital description may be needed to explain magnetic properties~\cite{Jeaong_2020} and to capture the proper pseudospin composition of the spectral function~\cite{Damascelli_2021}. 
Moreover, multi-orbital fingerprints can be observed via Hund's
coupling effects~\cite{Cassol_2024} and play a role in renormalizing
the electronic band gap~\cite{Pavarini_2023}.

Whether a quantitative and experimentally
accessible criterion can locate a given compound along the Slater-Heisenberg
crossover, and how that location evolves with $U$ in the 214 Ruddlesden-Popper iridates, are the
questions we address here. This is accomplished by
investigating the phase diagram of $t_{2g}^5$ iridates as a function of temperature $T$ and local interaction $U$ in a systematic way. 
Focusing on \sriro\ and \bairo, we cast both materials in a unified phase diagram. We find that both compounds are located in a region, where both itinerant and localized electrons cooperate to stabilize the antiferromagnetic insulator. 
We demonstrate that even though a single-band picture may capture the low-energy physics qualitatively, the \JEFF\ bands are not inactive spectators: their presence effectively weakens correlations and affects the stability of the magnetic phase, which results in a reduction of the Néel temperature at moderate coupling.
Remarkably, our phase diagram reconciles seemingly contradicting reports from previous literature.

%

%

\paragraph{Phase diagram of 214 Ruddlesden-Popper iridates.}
Figure~\ref{fig:Figure_1} summarizes the most salient features of the spin-orbit entangled 214 iridates as investigated in this work.
The $U-T$ phase diagram is obtained within DMFT~\cite{Georges_1996}, which treats local quantum fluctuations exactly within a self-consistent embedding scheme, describing on equal footing itinerant and localized electrons. 
Our reference is a three-orbital $t_{2g}$ Hubbard model with explicit SOC, derived for Ba$_2$IrO$_4$; the phase diagram of Sr$_2$IrO$_4$ is identical up to a rescaling of $U$ (see Extended Data Fig. 1). 
Throughout, $U$ denotes the local interaction acting on the \jeff\ manifold, so that it can be compared directly with the interaction of an effective single-band \jeff\ model, indicated by a dash-dotted line. 
The upper axis expresses $U$ in units of the \jeff\ bandwidth $W$.

In the three-orbital model, the phase diagram consists of three regions: a paramagnetic (PM) metal at weak coupling, a PM insulator at strong coupling, and an antiferromagnetic (AFM) insulator below the Néel temperature $T_N$ (solid black line). 
The AFM phase has the dome shape characteristic of the half-filled Hubbard model, as found in quantum Monte Carlo~\cite{Staudt_2000}, DMFT~\cite{Jarrell_1992}, and its cluster~\cite{Fratino_2017} and diagrammatic~\cite{Chatzieleftheriou_2024} extensions. 
$T_N$ rises steeply from about $150$ K at $U \simeq 1.3$ eV, reaches a maximum of about $800$ K at $U \simeq 2.1$ eV, and then decreases slowly at larger $U$. 
Since the ordered phase is insulating over the whole range of interactions, the nature of the transition at $T_N$ changes along the dome. 
At weak coupling, magnetic order turns a PM metal into an AFM insulator, i.e., the gap is opened by the order itself, typical of the Slater picture. 
At strong coupling, the electrons are already localized above $T_N$, and the transition only freezes the orientation of preformed local moments, characteristic of a Heisenberg antiferromagnet. 
The rise and fall of $T_N$ reflects this change.

\begin{figure}[tb]
    \centering
    \includegraphics[width=\linewidth]{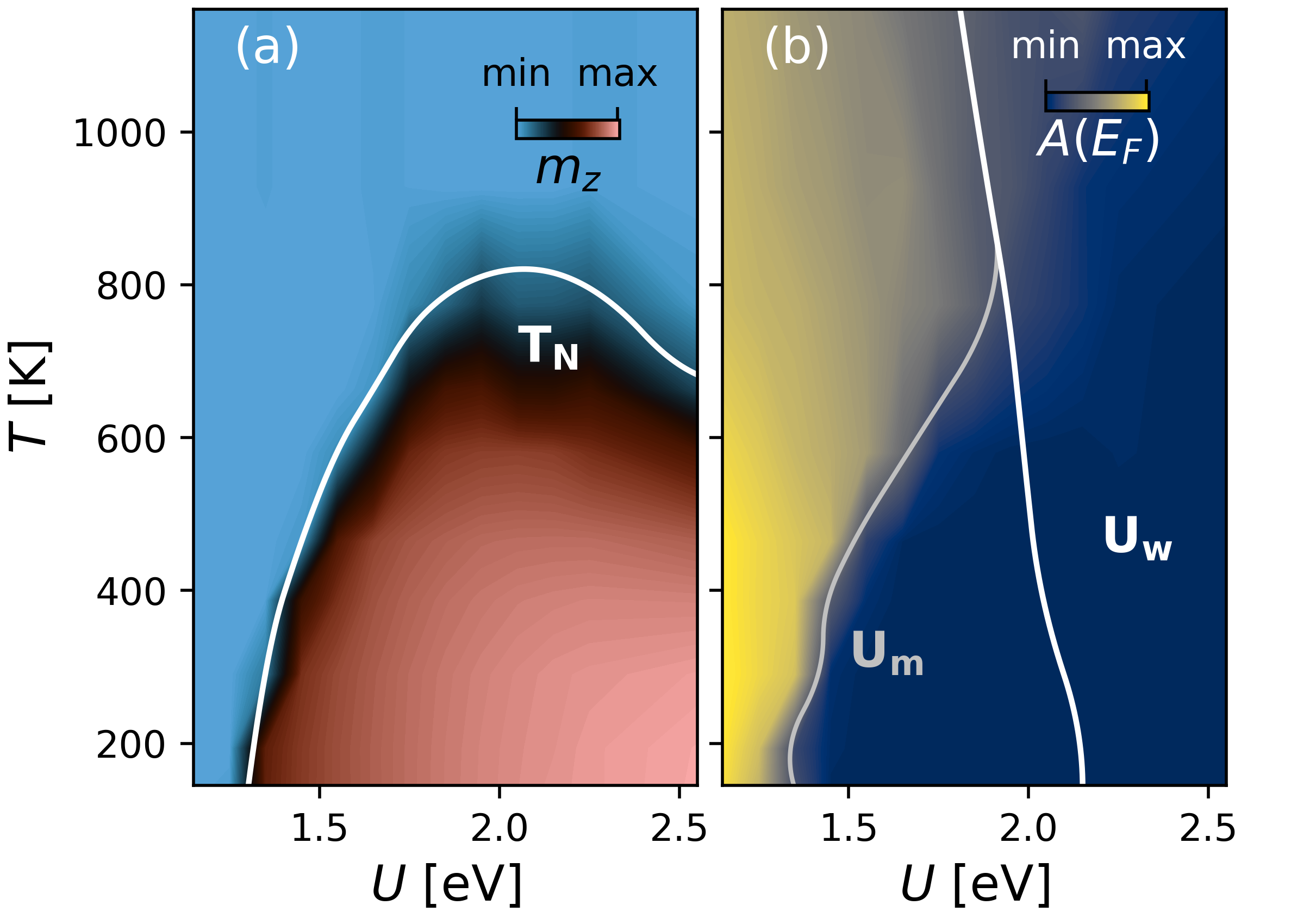}
    \caption{ Phase diagram details.
    Staggered magnetization $m_z$ (a) and spectral weight at the Fermi level $A(E_F)$ (b) as a function of temperature $T$ and interaction strength $U$.
    The magnetic transition temperature $T_N$ and the PM Widom line $U_W$ are indicated in white, whereas $U_m$ highlights the passage from metal to insulator due to AFM ordering.
    }
    \vspace{5pt}
    \hrule
    \label{fig:Figure_2}
\end{figure}

The two main phase diagram boundaries, $T_N$ and $U_W$, were extracted from the magnetic and the spectral properties of the system, respectively, as illustrated in Figure 2.
We used the order parameter $m_z=\frac{1}{2N}\sum_i e^{i\mathbf{Q}\cdot\mathbf{R_i}}(n_{i\uparrow}-n_{i\downarrow})$ to define $T_N$.
Here, $N$ is the total number of lattice sites, and the quantization axis is chosen to be in the $j_z$ direction, with $n_{i\sigma}$ being the occupation of the \jeff\ state with pseudospin $\sigma$ at site $i$.
Moreover, $\mathbf{Q}=(\pi,\pi)$ is the ordering wave vector, in line with the in-plane Néel pattern observed in \bairo~\cite{Okabe_2012,Isobe_2012}. 
Note that due to completely filled \JEFF\ bands  within the AFM phase, the magnetic moment develops in the \jeff\ sector only.
Since the \jeff\ state is a superposition of the $d_{xz}$,$d_{yz}$ and $d_{xy}$ orbitals, with weights set by the interplay between the tetragonal crystal field $\delta$ and the SOC strength $\lambda$~\cite{Cassol_2024}, all $t_{2g}$ states contribute to $m_z$, as shown in \ref{fig:Appendix_3}(a-c).
The low-temperature moment grows steadily with $U$.
It is small at  the weak-coupling edge of the dome, where it is controlled by the small Slater gap, and approaches saturation at large $U$, where it reflects fully formed local moments.
For each value of $U$, we define $T_N$ from fitting $m_z(T)$ according to the mean-field criticality expected from DMFT~\cite{Rohringer_2011}, see Methods and~\ref{fig:Appendix_4}.

The insulator is characterized by the suppression of spectral weight at the Fermi level, $A(E_F)$, and we define the metal-insulator line $U_m$ as the inflection point of $A(E_F)$ as a function of $U$ at fixed $T$. 
Below $T_N$, $U_m$ closely follows the weak-coupling side of the AFM dome: the opening of the gap coincides with the onset of magnetic order, as expected for a Slater insulator.
Above $T_N$, the PM phase evolves continuously from a metal to an insulator, as shown by the color gradient.
We locate this crossover through the Widom line $U_W$~\cite{Terletska_2011,Vucevic_2013}, which is obtained as the inflection point of $A(E_F)$ in the PM solution of the three-orbital model, i.e. when enforcing paramagnetism. 
$U_W$ is the continuation, above the critical endpoint, of the first-order Mott transition line, see \ref{fig:Appendix_4}(c), and it signals the charge localization by the local Coulomb repulsion. 
Inside the dome, it identifies the Mott crossover hidden beneath the magnetic order. 
$U_W$ shifts from $\simeq2.15$ eV at the lowest temperatures to lower values at higher temperatures, by emerging from the dome close to the maximum of $T_N$. 
This coincidence connects the non-monotonic behavior of $T_N$ to the underlying Mott physics: the ordering temperature is highest where the electrons of the PM phase lose their itinerant character.


\paragraph{Slater vs. Heisenberg antiferromagnetism.}

\begin{figure}[t!]
    \centering
    \includegraphics[width=\linewidth]{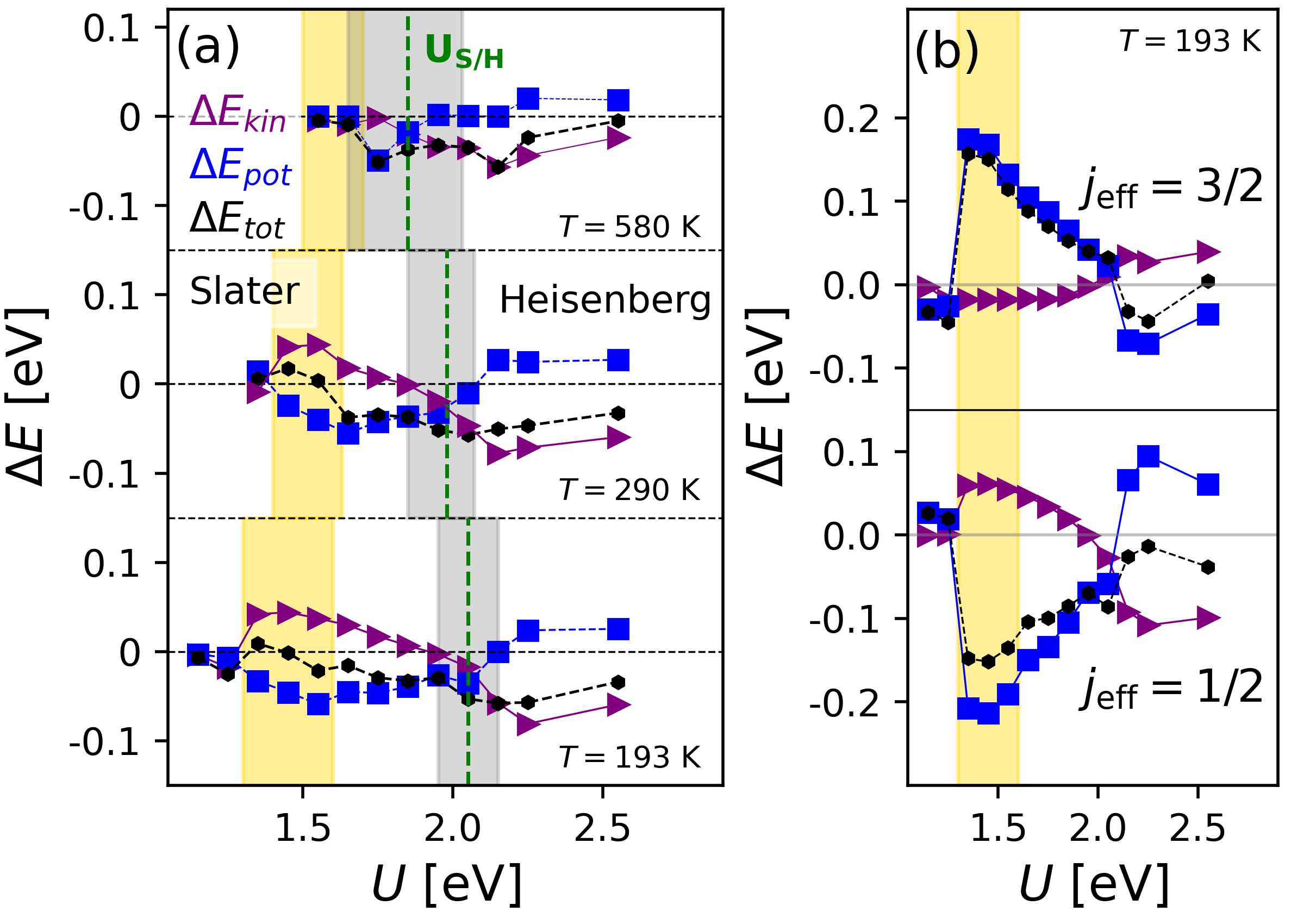}
    \caption{ Competition between AFM and PM phase. Energy difference between the AFM and the PM state, $\Delta E_{i} = E^{AFM}_{i}-E^{PM}_{i}$, in the kinetic, potential and total energy at different temperatures $T$. 
    The grey area indicates the passage from Slater to Heisenberg type of magnetism, the green dashed lines denote the middle of this crossover $U_{S/H}$, consistent with  \figurename~\ref{fig:Figure_1}. The coexistence region is indicated as a yellow area.
    Panel (b) shows the energy difference projected into the \jeff\ and \JEFF\ subspaces. 
    }
    \vspace{5pt}
    \hrule
    \label{fig:Figure_3}
  \end{figure}

The character of the magnetic order is quantified by the line $U_{S/H}$, which marks the passage from Slater to Heisenberg magnetism. 
This attribution is based on the energy differences between the AFM and PM phases, $\Delta E=E^{\mathrm{AFM}}-E^{\mathrm{PM}}$, separated into the kinetic and potential contributions, see Fig.~\ref{fig:Figure_3}.
Overall, the situation resembles that of the half-filled single-band Hubbard model: The magnetic ordering is driven by potential energy gain in the Slater regime, and by kinetic energy on the Heisenberg side~\cite{Taranto_2012,Fratino_2017,Crispino_2023}.
The point where the dominating contribution switches, $\Delta E_{pot} = \Delta E_{kin}$, defines $U_{S/H}$. 
Within a certain range around $U_{S/H}$, itinerant and localized electrons cooperate in stabilizing the AFM phase, which is therefore of mixed type.
The AFM dome is thus divided into three regimes: a Slater regime for small $U$, a cooperative regime around $U_{S/H}$ where reduction in both kinetic and potential energy with respect to the PM solution stabilizes the magnetic order, and a pure Heisenberg regime for larger $U$. 
Notice how $U_{S/H} \simeq 1.8-2.0$ eV lies systematically below $U_W$. 
Just above $U_{S/H}$ the contribution of localized electrons becomes dominant. 
Even in the intermediate regime, magnetism is already influenced by electron localization, although the PM solution at the same $U$ is still on the metallic, itinerant side of the Mott crossover. 
Thus, magnetic order promotes the localization of the \jeff\ electrons.
Whereas the presence of the cooperation region is not a general phenomenon in correlated materials with underlying multi-band Mott transition~\cite{Crispino_2026}, it is a robust feature in the 214 iridates and is also found when employing techniques including non-local fluctuations~\cite{Watanabe_2014}.

\paragraph{Single- vs. Multi-band character.}

\begin{figure}[t]
\centering
    \includegraphics[width=\linewidth]{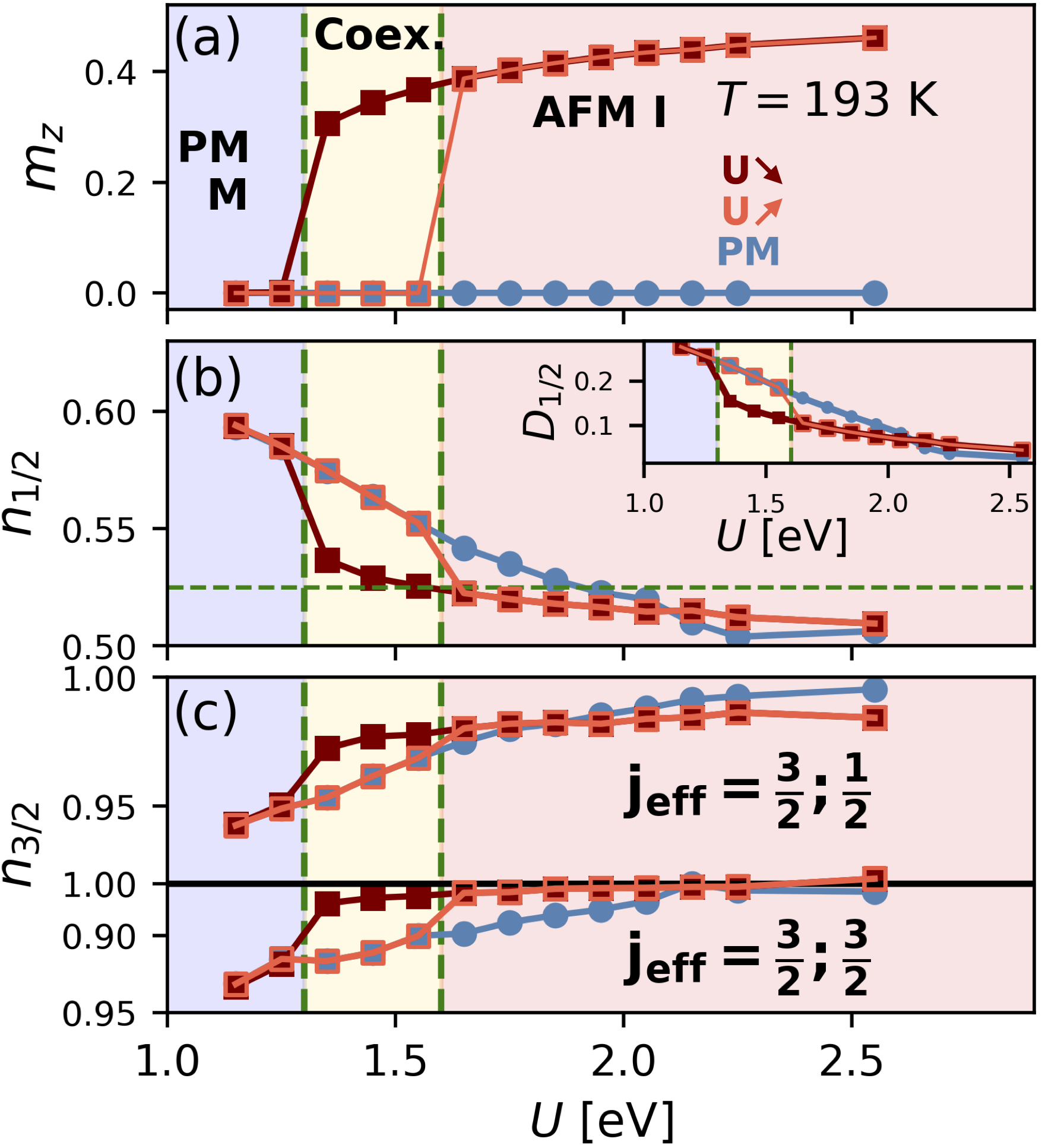}
    \caption{ Magnetic transition and coexistence region. Staggered magnetization (a), filling of the \jeff\ (b) and \JEFF\ states (c) respectively as a function of interaction strength. 
    The inset shows the double occupancy $D_{1/2}$ of the \jeff\ states.
    The 
    U$\shortarrow{1}$ (U$\shortarrow{7}$)
    line indicates the solution obtained by 
    increasing (decreasing) $U$; the constrained PM solution is shown in blue.
    }
    \label{fig:Figure_4}
    \vspace{5pt}
    \hrule
  \end{figure}

At the weak-to-intermediate coupling edge of the dome, a coexistence region extends from the $T_N$ line up to $U \simeq 1.6-1.7$ eV.
There, stable PM and AFM solutions coexist, and following a given solution within DMFT while decreasing/increasing $U$ leads to hysteretic behavior characteristic of a first order transition, see Fig.~\ref{fig:Figure_4}.
This coexistence stems from the competition between the three-orbital PM metal and the AFM insulator, which involves charge transfer between the \jeff\ and \JEFF\ manifolds.

The comparison with the single-band \jeff\ model (see Fig.~\ref{fig:Figure_1}) illustrates the multi-orbital regime of the iridates. 
The single-band model gives a qualitatively similar dome, with the same sequence of regimes as a function of $U$. 
Quantitatively, however, the two models differ in three aspects. 
First, the three-orbital $T_N$ is strongly suppressed at weaker coupling, whereas the two curves converge at strong coupling. 
This is understood by the fact that in the Heisenberg regime the \JEFF\ bands are inert and the exchange is set by the \jeff\ hopping alone. 
Second, the maximum of $T_N$ moves to larger $U$ 
in the three-orbital model, and both $U_W$ and $U_{S/H}$ are likewise found at smaller $U$ in the single-band case (\ref{fig:Appendix_2}). 
The \JEFF\ bands thus screen the interaction within the \jeff\ manifold, so that a larger bare $U$ is needed to reach the same degree of localization. 
Third, the coexistence region found here is clearly a multi-band effect, at variance with the single-band model, where hysteresis at intermediate coupling can come from lattice frustration~\cite{Raczkowski_2021}. 
The single-band description is therefore applicable on the strong-coupling side, but it misses the multi-orbital physics precisely in the weak and intermediate coupling regime.

To clarify the multi-band interplay, $\Delta E_{kin}$ and $\Delta E_{pot}$ are projected on the two $j_{\mathrm{eff}}$ sub-manifolds, as shown in \figurename~\ref{fig:Figure_3}(b). 
The onset of the magnetic phase is accompanied by a jump in potential energy of the \JEFF\ bands. Whilst the \jeff\ contribution stabilizes the magnetic state, the \JEFF\ bands favor the paramagnetic metal.
This competition is also reflected in the evolution of the orbital filling, see \figurename~\ref{fig:Figure_4}.
At weak interaction, a small but finite amount of holes is present within the \JEFF\ bands, which thus cross the Fermi level, similarly to the situation within density functional theory (DFT) calculations~\cite{Martins_2011,Cassol_2024}.
Both the \jeff\ and \JEFF\ densities show a jump at the transition to the AFM insulator, i.e. at the boundaries of the coexistence region, within which both an AFM insulator and a PM metal can be stabilized. 

Increasing local interactions (U$\shortarrow{1}$ in \figurename~\ref{fig:Figure_4}), the change in \JEFF\ densities at the transition is rather small, due to an effective correlation-induced enhancement of the SOC magnitude, 
as previously reported for rhodates and ruthenates~\cite{Zhang_2016}. 
However, when decreasing $U$ along the U$\shortarrow{7}$ side of the hysteresis, the change in $n_{1/2},n_{3/2}$ is more important. 
Note that the double occupancy of the \jeff\ pseudospin $D_{1/2}$ mirrors the evolution of $n_{1/2}$. 
Thereby, the transition to a single, half-filled \jeff\ band is always accompanied by the magnetic metal insulator transition.

The charge transfer from the \jeff\ to the \JEFF\ manifold at the transition was previously noted in the literature, but limited to the PM phase~\cite{Martins_2011,Cassol_2024}, where a Lifshitz transition (LT)~\cite{Lifshitz} was detected at moderate $U$, modifying the topology of the Fermi surface in the metallic regime~\cite{Cassol_2024}. 
Here, the LT coincides with the AFM gap opening, preventing a direct observation of this effect in the metallic phase.
Since the \JEFF\ bands experience larger local interaction (see \ref{tab:Tab_2}), essentially due to the form of the SOC Hamiltonian~\cite{Martins_2017}, the abrupt charge transfer from the \jeff\ bands amounts to an increase in potential energy. 
At stronger interaction, the energetic cost linked to the charge transfer is reduced and favors the AFM phase.

\begin{figure*}[h]
\centering

    \includegraphics[width=\linewidth]{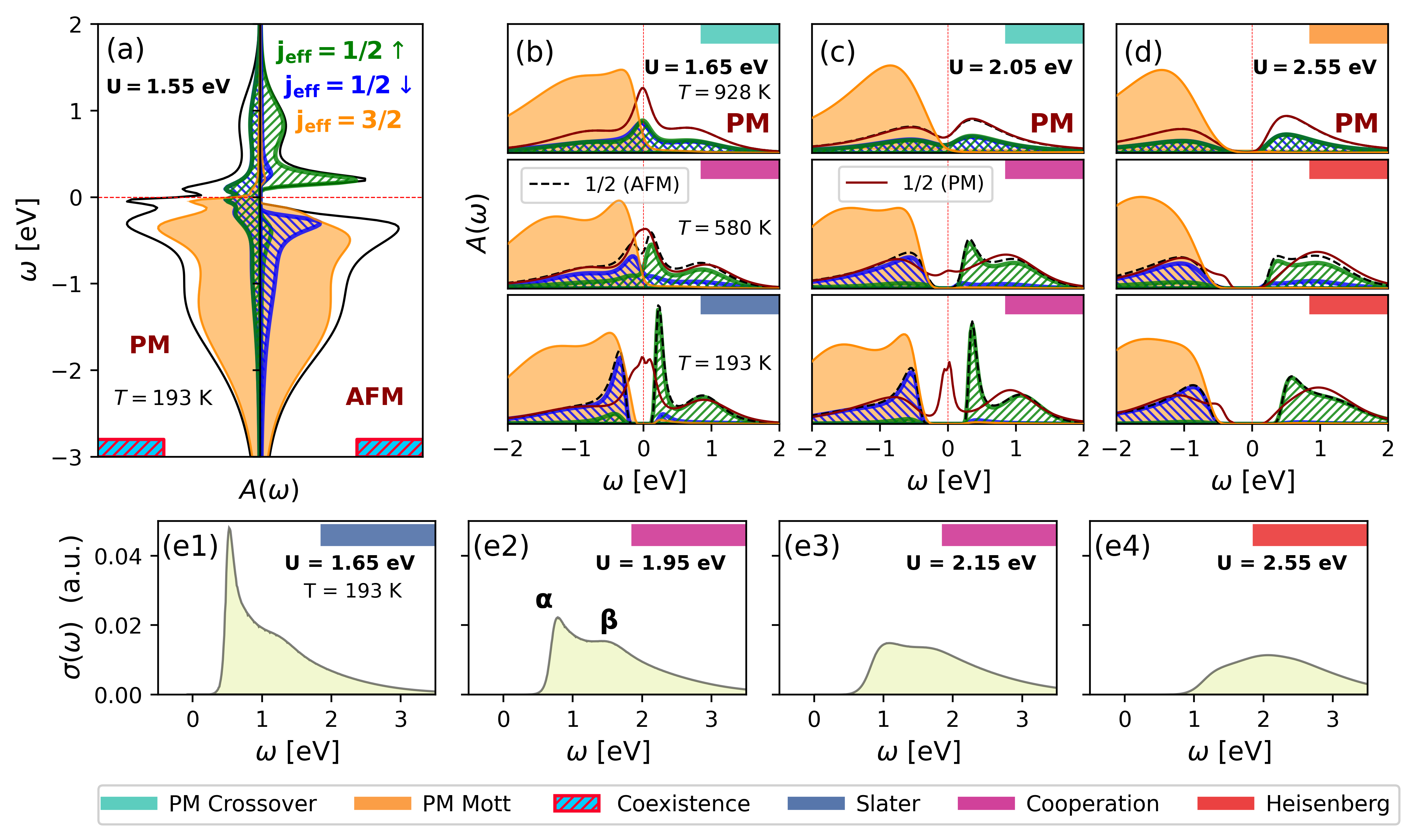}
    \caption{ 
     Evolution of the local spectral function.
    (a) Comparison of the PM metal and AFM insulating solution in the coexistence region, highlighting the different orbital contributions. (b-d) Temperature dependence of the local spectral function for electron-electron interaction strengths situating the system in different regions of the phase diagram of Fig.~\ref{fig:Figure_1}.
    The total \jeff\ contribution is indicated by a dashed line and, for comparison, we also indicate the \jeff\ component in the PM phase by solid brown lines in order to show the approximate coincidence with the Hubbard sub-bands observed in the magnetic state. 
    (e1-e4) Optical conductivity at $T=193\ $K, showing a redistribution of weight between $\alpha$ and $\beta$ peak with increasing $U$.
    }
     \vspace{5pt}
    \hrule
    \label{fig:Figure_5}
\end{figure*}

\paragraph{Spectral properties.}

The transfer of spectral weight is best seen in the local spectral function $A(\omega)$ within the coexistence region, see Fig.~\ref{fig:Figure_5}(a).
Finite spectral weight of \JEFF\ character remains at the Fermi level in the PM phase, contrasting with the AFM insulator where the unoccupied part of the spectrum above the spectral gap has nearly exclusive \jeff\ character. 
In that phase, the characteristic spin-polaron quasiparticle peaks close to the gap are well pronounced and separated from the incoherent part of the upper and lower Hubbard subbands.

Based on three representative $U$ values close to the maximum of the magnetic dome, we show in \figurename~\ref{fig:Figure_5}(b-d) the evolution of the local spectral function $A(\omega)$ with temperature.
It is in this region that most studies in literature have located \bairo\ and \sriro.
The PM high-temperature phase shows the metal-insulator crossover, while in the low-temperature phase the energetics allow to identify the AFM phase at $U=1.65\ $eV as Slater-type, at $U=2.55\ $eV as Mott-Heisenberg like and as cooperation regime for $U=2.05\ $eV.

Below $T_N$, the spectral function opens a gap and develops double-peak structures, each composed of a spin-polaron (SP) quasiparticle peak, responsible for shaping the optical spectrum~\cite{Cassol_2026}, and the Hubbard band.
These characteristics of the iridates' \jeff\ band structure are not only present in AF-DMFT studies~\cite{Arita_2012,Zhang_2013} but also within PM cluster DMFT model calculations~\cite{Fratino_2017}. 
In the latter, the electronic properties arise from short-range non-local fluctuations, which prevail even in the PM phase close to the magnetic phase transition and do not lead to any notable change in $A(\omega)$ across the transition.
Within AF-DMFT non-local fluctuations are absent, but their effects are mimicked by static antiferromagnetism.
The incoherent Hubbard subbands are also found when artificially enforcing a PM solution, suggesting its interpretation as a "Mott-like" spectral feature.
However, the separation of the SP and the Hubbard subbands gets washed out when increasing $U$:
Spectral weight is transferred from the spin-polaron to the Hubbard bands, merging the two. 
On the other hand, decreasing the temperature renders the SP more coherent. 
Overall, the changes in the spectral function remain quite moderate below $T_N$ within the cooperation region. 

Consequences of this spectral weight transfer are much more evident in the optical conductivity $\sigma(\omega)$, see \figurename~\ref{fig:Figure_5}(e1-e4).
Since the coherent SP peak is necessary to shape the $\alpha$ peak in $\sigma(\omega)$, whereas transitions to the upper Hubbard sub-band contribute mainly to the $\beta$ peak~\cite{Cassol_2026}, neither the Slater- nor the Heisenberg regime can reproduce its characteristic two-peak structure. 
Only in the cooperation region balanced contributions to both peaks are obtained.

Focusing instead on the behavior close to the magnetic transition only allows to draw a clear picture deep in the Mott and Slater regimes (see e.g. Fig.~\ref{fig:Figure_5}(b,d)), when checking for a closing of or a significant change in the spectral gap at $T_N$.
However, for $U=2.05$ eV, the situation is far from clear since the material is in a bad insulator phase at $T\gtrsim T_N$, but develops well defined SP quasiparticle peaks at low temperature.
It is precisely in this region that a clear attribution of the nature of the insulating phase to spectral features is problematic.

This shows that in the cooperation region neither the presence of SP features in the spectral function at low-temperature, nor its behavior across the magnetic transition temperature can possibly serve as an unambiguous proxy for the nature of its low-temperature insulating state.
In contrast, the balanced contribution of $\alpha$ and $\beta$ peaks in the optical conductivity clearly indicates the relevance of this parameter regime for the iridates.

\paragraph{A unified perspective.}
Drawing a link between correlations and magnetic properties of the 214 Ruddlesden-Popper iridates is possible via the phase diagram. 
Locating the materials within it by attributing a correct $U$ value is, however, crucial. 

Estimates of $U$ can be based on comparison to experimental probes or theoretical \textit{ab initio} techniques.
The former lead to a broad range of $U$ values: 
Angle-resolved photoemission spectroscopy (ARPES) band structures are well reproduced around the maximum of the AFM dome, see \ref{fig:Appendix_5}(c,d).
The underlying weak dependence of the \jeff\ dispersion on $U$ explains the large range of values successfully employed previously in literature~\cite{Zhang_2013,Lenz_2019, Choi_2024, Cassol_2026}, see~\ref{tab:Tab_1}.
However, bulk-sensitive probes like optical conductivity~\cite{Seo_2017} lead to smaller optical gaps than local probes like ARPES~\cite{Brouet_2015} or scanning tunneling spectroscopy (STS)~\cite{Dai_2014}, pointing to a problem \textit{per se}. 
In turn, tuning $U$ to match these experimental gaps would lead to a change in the underlying physics:
Whereas the gap of optical conductivity ($\sim150\ \mathrm{meV}$) locates the materials in the Slater-Heisenberg crossover, the STS and ARPES gap ($\sim700\ \mathrm{meV}$) would place them deep in the Mott-Heisenberg region.
Apart from the latter, most experiments would place both materials in the crossover regime, where Slater and Mott physics coexist.  
There, the N\'eel temperature does not vary strongly with $U$ around the maximum of the AFM dome. 
The values reported for the two compounds~\cite{Kim_2008,Okabe_2012} can therefore correspond to different positions within the crossover region.

\begin{figure}[tb]
\centering
    \includegraphics[width=\linewidth]{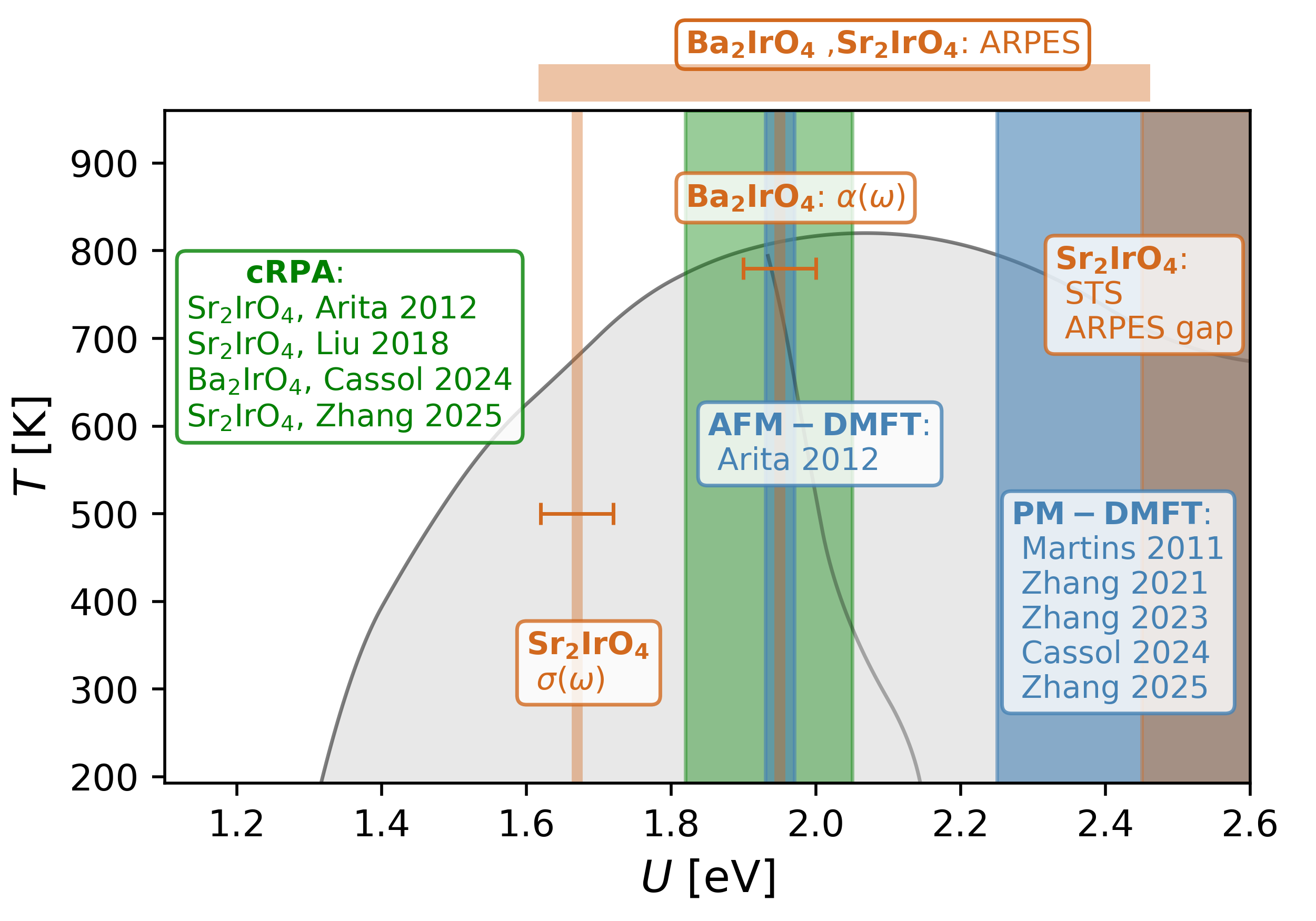}
    \caption{Locating \bairo\ and \sriro\ in the phase diagram. Black lines represent the Néel temperature and the Widom line.
    Shaded orange areas denote the values of the Hubbard $U$ to be set to reproduce experimental observables, such as transport quantities~\cite{Nichols_2014,Seo_2017} or electronic gap from  local probes. Blue regions highlight values employed in previous DMFT studies~\cite{Martins_2011,Arita_2012,Pavarini_2021,Pavarini_2023,Cassol_2024,Pavarini_2025}. The green range contains values reported from previous cRPA calculations~\cite{Arita_2012,Liu_2018,Cassol_2024,Pavarini_2025}. All numerical values are reported in \ref{tab:Tab_1}. 
    }
    \vspace{5pt}
    \hrule
    \label{fig:Figure_6}
\end{figure}

\textit{Ab initio} calculations of the interaction based on cRPA lead for both materials to $U$ values close to the maximum of the AFM dome~\cite{Arita_2012,Cassol_2024,Pavarini_2025}. 
However, given the tendency of cRPA to overscreen and the neglect of the dynamical part of the interaction within DMFT, these values have to be seen as rough estimates of a suitable static local $U$ value.

Comparing the interaction strengths used in earlier studies with our phase diagram helps to reconcile their seemingly conflicting conclusions.
In Refs.~\cite{Martins_2011,Pavarini_2023,Cassol_2024,Pavarini_2025}, the authors have focused on the PM phase, and have employed interaction values larger than the cRPA reference, corroborating a Mott picture. 
In \figurename~\ref{fig:Figure_6}, all these values are indeed located beyond the Widom line. 
Closer examination is needed when comparing to Arita \textit{et al.}~\cite{Arita_2012}, where the authors employ cRPA with a dynamical exchange-correlation kernel and support a Slater picture based on the closure of the electronic gap above $T_N$. 
Albeit featuring a metallic solution above $T_N$, this value is close to the Widom line, as shown in \figurename~\ref{fig:Figure_6}.
This further confirms the importance of a detailed investigation of the system's phase diagram and of its energetics, in order to investigate the origin of the antiferromagnetic insulating phase. 

A better consistency between the theoretically predicted and the experimentally tailored $U$ values could be achieved by the explicit inclusion of $e_g$ states or oxygen ligand orbitals, as recently shown in the context of cuprates~\cite{Vucicevic_2026}. 
Also the description of the Widom line separating itinerant from localized physics could conceptually be improved by including non-local extensions of DMFT, with non-trivial consequences on its precise location~\cite{Fratino_2017,Chatzieleftheriou_2024}.
However, a rigid classification of \bairo\ and \sriro\ as either Slater or Mott insulators based on DMFT seems to be incomplete.
Based on the evidence shown here and consistent with the experimental finding of both Slater and Mott fingerprints~\cite{Hsieh_2012}, they are prototypical examples of materials where localized and itinerant physics cooperate to form the antiferromagnetic insulating low-temperature phase.
Providing a complete overview of their physics, the present phase diagram can thereby constitute a compass for future studies on the 214 Ruddlesden-Popper iridates and other spin-orbit coupled correlated materials.

\section*{Methods}

\paragraph{Computational details.}

Electronic correlations are taken into account employing density functional theory (DFT) together with dynamical mean-field theory (DMFT)~\cite{Georges_1996} of correlated electrons. 

The density functional theory (DFT) calculations were performed using plane wave basis sets within the Quantum Espresso code~\cite{QuantumEspresso1}. 
Following the procedure outlined in~\cite{Cassol_2024,Cassol_2026}, norm-conserving 
pseudopotentials from the Pseudodojo library~\cite{PseudoDojo2018} were employed.
The three-band tight-binding model for \bairo\ was obtained wannierizing non-relativistic calculations and adding SOC as a local term.
The single-band model was obtained by wannierizing the spin-polarized fully relativistic band structure, in an energy window that only contains the \jeff\ bands, see next section.

In the case of \sriro, a three-orbital 2D in-plane model was constructed wannierizing the DFT band structure directly, like in Ref.~\cite{Cassol_2026}.

Dynamical mean-field theory (DMFT) calculations were performed with the TRIQS library \cite{TRIQS2015} and 
the DMFT impurity problem solved within the Continuous time quantum Monte Carlo (CTQMC) solver in Hybridization expansion variant (CT-HYB) \cite{CTQMC}. 
The DMFT self-consistency loop was coded taking advantage of the DFTTools module \cite{TRIQSDFTTOOLS2016}, and the impurity problem was solved
with $28 \times 10^6$ Monte Carlo measurements per iteration. 
To obtain a reliable analytic continuation, we increased the statistics to $86 \times 10^6$ measurements for the final iteration. 
Magnetism was accounted for in a mean-field manner, thereby assuming static magnetic order. This is obtained by partitioning the original system into two sublattices $A$ and $B$, hybridizing through the bath. 
The corresponding impurity model was solved for one of the two sublattices. 
The self-energy of the second sublattice, $\Sigma_B(i\omega_n)$, was obtained by applying time reversal symmetry to $\Sigma_A(i\omega_n)$ in the $j_{\mathrm{eff}}$ basis. 
For most of the phase diagram, the systems spontaneously evolved to a more stable configuration independently of the starting $\Sigma(i\omega_n)$. 
However, within the coexistence region, a starting paramagnetic (antiferromagnetic) $\Sigma(i\omega_n)$ was needed to converge the paramagnetic (antiferromagnetic) solution. 
In particular, the paramagnetic solutions were stabilized converging a spin-average PM solution and subsequently releasing the constraint.  To identify the upper  $U$ boundary, we considered stable the solutions that, after performing 10 iterations, fulfilled $max|G_n(\tau)-G_{n-1}(\tau)|< 10^{-3}$ and $m_z < 10^{-3}$. Due to fermionic sign problem,  larger precision is  unfortunately not affordable. Hence, the right boundary of the coexistence region should be regarded as a guide to the eye.
Within the rest of the magnetic dome, paramagnetic calculations needed to define the Widom line and to inspect the energy balance were simply stabilized imposing an equal pseudospin occupation. 

The analytical continuation was performed with the matrix generalization of the maximum entropy method as implemented in the TRIQS toolkit \cite{MEM_1}.

All calculations presented in the main text are performed with the \textit{ab initio} electronic structure of \bairo, but they are representative of \sriro\ as well, as shown in \ref{fig:Appendix_1}.

\paragraph{Hubbard models and spin-orbit interaction.}

Starting from the DFT reference, the lattice Hamiltonian is written in the $t_{2g}$ subspace as the sum of three terms: 

\begin{equation}
    \widehat{H} = \widehat{H}_{kin} + \widehat{H}_{soc} + \widehat{H}_{int} \ ,
\end{equation}

where the first term represents the kinetic energy in the $t_{2g}$ basis and involves the intersite and interorbital hoppings. 
As reported in Ref. \cite{Cassol_2024}, the system presents some tetragonal splitting $\delta$, 
which is caused by the ligand field, 
since the apical oxygen atoms participate in the Ir-O-Ba bonds while the basal oxygen atoms contribute to Ir-O-Ir. 

The SOC is incorporated as a local term in the Hamiltonian, parametrized by the SOC constant $\lambda$ \cite{Cassol_2024,Gaspard_2026}: 

    \begin{equation}
            \label{eq:model_SOC}
         \widehat{H}_{\text{SOC}} =  \lambda \sum_{\alpha} {\hat{\mathbf{L}}}_\alpha \cdot {\hat{\mathbf{S}}}_\alpha \ .
    \end{equation} 

The value of $\lambda$ is chosen to fit the band structure obtained within DFT+SOC  
leading  to a value of $\lambda = 0.31$ eV (see \ref{fig:Appendix_7}(a)).
Notice that in the context of the single-band \jeff\ model the wannierization was performed on a single band only. Within the three-orbital model, the magnitude of $\lambda$ 
defines the splitting between  the \jeff\ and the \JEFF\ submanifolds, as it is schematically depicted in panel (b).

Without the effect of the tetragonal distortion $\delta$, all $t_{2g}$ orbitals would contribute with the same weight of 1/3 to the \jeff\ state. 
However, for non-vanishing $\delta$, one has a deviation from this cubic symmetry situation, as it was already pointed out in Refs.~\cite{Martins_2011,Cassol_2024}. 
In our case, the presence of a tetragonal ligand field yields a lower contribution of the $d_{xy}$ orbital to the \jeff\ pseudospin, which displays larger $d_{xz}/d_{yz}$ character. 
Conversely, the $d_{xy}$ character is prominent in the $\ket{3/2;\pm 1/2}$ state, while $\ket{3/2;\pm 3/2}$ remains of pure $d_{xz}/d_{yz}$ character. 
    
The Coulomb interaction is expressed in the $t_{2g}$ basis in the generalized Hubbard Kanamori form: 

\begin{equation}
\begin{split}
\hat{H}_{\mathrm{int}} =
& \frac{1}{2} \sum_{\sigma} \sum_{ij} U_{ij}\, \hat{n}_{i\sigma}\hat{n}_{j\bar{\sigma}} 
+ \frac{1}{2} \sum_{\sigma} \sum_{i \neq j} (U_{ij} - J_{ij})\, \hat{n}_{i\sigma}\hat{n}_{j\sigma} \\
& - \frac{1}{2} \sum_{\sigma} \sum_{i \neq j} J_{ij}
\left[
\cdag_{i\sigma}\c_{i\bar{\sigma}}\cdag_{j\bar{\sigma}}\c_{j\sigma}
-
\cdag_{i\sigma}\cdag_{i\bar{\sigma}}\c_{j\sigma}\c_{j\bar{\sigma}}
\right]
\end{split}
\label{eq:HK}
\end{equation}

composed by density-density, spin-flipping, and pair hopping terms, respectively. 

This results in a small anisotropy in the Coulomb tensor, which displays enhanced interactions for the $d_{xy}$ orbital. 
The full Coulomb tensor was previously evaluated from cRPA in Ref. \cite{Cassol_2024}. 
We report here the values for completeness. 
In order to span the phase space, we add to the density-density terms a constant amount $\Delta U$. 
This is meant to account for uncertainties on the screening coming
from the cRPA technique, which mainly affects the $F_0$ Slater
integral.

Due to fermionic sign problem, all calculations must be performed in the \jeff\ basis and the Coulomb interaction must be rotated accordingly. 
The resulting Coulomb tensor thereby contains additional elements not
included in the interaction form of equation~\ref{eq:HK}, such as correlated-hopping and four-scattering terms \cite{Kim_2020}. 
The density-density and exchange terms of the rotated cRPA Coulomb tensor are reported in \ref{tab:Tab_2}.

\paragraph{Total energy.}

In order to determine the energy balance of the system standard formulas were employed. The kinetic term was evaluated as: 

\begin{equation}
    E_{kin} =  \text{Tr} \left[ H_{-\mathbf{k}}G_{\mathbf{k}}\right] =\frac{1}{N_{\mathbf{k}}\beta}\sum_{\mathbf{k}}\sum_{i\omega_n} \text{Tr} \left[ H_{-\mathbf{k}}G_{\mathbf{k}}(i\omega_n) \right]
\end{equation}

where $G_{\mathbf{k}}(i\omega_n)$ is the lattice Green's function and $H_{\mathbf{k}}$ is the Kohn Sham Hamiltonian.

The potential energy was obtained employing the Galitskii-Migdal formula: 

\begin{equation}
    E_{pot} =  \frac{1}{2\beta}\text{Tr} \left[ \Sigma G \right] =\frac{1}{2\beta} \sum_{i\omega_n} \text{Tr} \left[ \Sigma(i\omega_n) G(i\omega_n) \right]
\end{equation}

which is consistent with taking the statistical average $\braket{\widehat{H}_{int}}$. 
In order to extract the orbital components, we truncated the trace in the orbital space after  momentum and Matsubara sums.

\paragraph{Fitting procedures.}

To interpolate $m_z$ and $A_{E_F}$, and determine the transition lines, two schemes were employed. 
To study the metal-insulator transition,  $A_{E_F}$ was defined on a coarse grid $(U, T)$. 
Subsequently, the evolution of its \jeff\ component at fixed $T$ was interpolated with the following function

   \begin{equation}
       A^{1/2}_{E_F}(U)=\frac{a(T)}{1+e^{b(T)(U - U_W(T))}} \ ,
       \label{eq:fit_G_b_2}
   \end{equation}
   
 as it was done in Ref.~\cite{Cassol_2024} to study the Mott transition as a function of $\lambda$. 
 The choice of the \jeff\ component is justified since the metal-insulator transition occurs in this manifold.
 $U_W(T)$ represents the middle of the crossover region, and defines the Widom line. 
 Albeit not being a physical parameter, $b(T)$ intuitively defines the crossover's size. 
Once $a,b$ and $U_W$ are determined for all temperatures, we define them on a denser $T$ grid via a polynomial interpolation. 

For given $T$, the beginning and the end of the crossover are defined considering the extrema of the third derivative of the Fermi function, determined analytically by 

\begin{equation}
    U_{>/<}(T) = U_W(T) + \frac{1}{b(T)}\log\left(\frac{ 1/2 \pm 1/\sqrt6}{1/2 \mp 1/\sqrt6}\right) \ .
\end{equation}

On the other hand, the temperature evolution of the staggered magnetization $m_z$ was fitted with the function:

\begin{equation}
     m_z (U)= A(U)(T_N(U)-T)^{\frac{1}{2}},
\end{equation}

which is a standard power law using the mean-field (Gaussian) critical exponent $\beta=1/2$ \cite{Rohringer_2011}, and determines the Néel temperature $T_N(U)$.

Examples of fitting procedures are reported in \ref{fig:Appendix_4}.

\section{Acknowledgments}
The authors thank Silke Biermann, Véronique Brouet, Christophe Brun, Massimo Capone, Léo Gaspard, Cyril Martins,  Luca de' Medici, Simon Moser, Giorgio Sangiovanni, and Evgeny Stepanov for fruitful discussions. 
 This project was provided with HPC computing and storage resources by GENCI at TGCC thanks to the grant A0190912043 on the supercomputer Joliot Curie (ROME partition).

\section{Competing Interests}
The authors declare that they have no conflicts of interest that could influence the results or interpretation of this work.

\section{Data availability}

All data supporting the findings of this study are available from the authors upon reasonable request. 

\bibliographystyle{apsrev4-2}

\bibliography{biblio}

@misc{Gaspard_2026,
    title = {{Affordable Five-Orbital Dynamical Mean-Field Theory for Layered Iridates and Rhodates}},
    author = {Gaspard, L{\'e}o and Martins, Cyril},
    year = {2026},
    eprint = {2601.04832},
    archivePrefix = {arXiv},
    primaryClass = {cond-mat.str-el},
    url = {https://arxiv.org/abs/2601.04832},
}

@article{PseudoDojo2018,
  title = {{The {{PseudoDojo}}: {{Training}} and Grading a 85 Element Optimized Norm-Conserving Pseudopotential Table}},
  shorttitle = {The {{PseudoDojo}}},
  author = {{van Setten}, M. J. and Giantomassi, M. and Bousquet, E. and Verstraete, M. J. and Hamann, D. R. and Gonze, X. and Rignanese, G. -M.},
  year = {2018},
  month = may,
  journal = {Computer Physics Communications},
  volume = {226},
  pages = {39--54},
  issn = {0010-4655},
  doi = {https://doi.org/10.1016/j.cpc.2018.01.012},
  urldate = {2023-05-09},
  langid = {english},
}

@article{TRIQSDFTTOOLS2016,
  title = {{{{TRIQS}}/{{DFTTools}}: {{A TRIQS}} Application for Ab Initio Calculations of Correlated Materials}},
  shorttitle = {{{TRIQS}}/{{DFTTools}}},
  author = {Aichhorn, Markus and Pourovskii, Leonid and Seth, Priyanka and Vildosola, Veronica and Zingl, Manuel and Peil, Oleg E. and Deng, Xiaoyu and Mravlje, Jernej and Kraberger, Gernot J. and Martins, Cyril and Ferrero, Michel and Parcollet, Olivier},
  year = {2016},
  month = jul,
  journal = {Computer Physics Communications},
  volume = {204},
  pages = {200--208},
  issn = {0010-4655},
  doi = {https://doi.org/10.1016/j.cpc.2016.03.014},
  urldate = {2023-05-09},
  langid = {english},
}

@article{Seo_2017,
	title = {{Infrared probe of pseudogap in electron-doped $\text{Sr}_2 \text{IrO}_4$}},
	volume = {7},
	copyright = {2017 The Author(s)},
	issn = {2045-2322},
	url = {https://www.nature.com/articles/s41598-017-10725-z},
	doi = {10.1038/s41598-017-10725-z},
	number = {1},
	urldate = {2025-02-17},
	journal = {Scientific Reports},
	author = {Seo, J. H. and Ahn, G. H. and Song, S. J. and Chen, X. and Wilson, S. D. and Moon, S. J.},
	month = sep,
	year = {2017},
	pages = {10494},
}

@article{moon_2009,
	title = {{Temperature dependence of the electronic structure of the ${J}_{\text{eff}}=1/2$ Mott insulator $\text{Sr}_{2}\text{IrO}_{4}$ studied by optical spectroscopy studied by optical spectroscopy}},
	volume = {80},
	url = {https://link.aps.org/doi/10.1103/PhysRevB.80.195110},
	doi = {10.1103/PhysRevB.80.195110},
	number = {19},
	urldate = {2025-02-17},
	journal = {Physical Review B},
	author = {Moon, S. J. and Jin, Hosub and Choi, W. S. and Lee, J. S. and Seo, S. S. A. and Yu, J. and Cao, G. and Noh, T. W. and Lee, Y. S.},
	month = nov,
	year = {2009},
	note = {Publisher: American Physical Society},
	pages = {195110},
}

@article{Martins_2011,
  title = {{Reduced Effective Spin-Orbital Degeneracy and Spin-Orbital Ordering in Paramagnetic Transition-Metal Oxides: ${\mathrm{Sr}}_{2}{\mathrm{IrO}}_{4}$ versus $\text{Sr}_2 \text{RhO}_4$}},
  author = {Martins, Cyril and Aichhorn, Markus and Vaugier, Lo\"{\i}g and Biermann, Silke},
  journal = {Phys. Rev. Lett.},
  volume = {107},
  issue = {26},
  pages = {266404},
  numpages = {5},
  year = {2011},
  month = {Dec},
  publisher = {American Physical Society},
  doi = {10.1103/PhysRevLett.107.266404},
  url = {https://link.aps.org/doi/10.1103/PhysRevLett.107.266404}
}

@article{Cassol_2024,
  title = {{Rich phase diagram of the prototypical iridate $\text{Ba}_2 \text{IrO}_4$: Effective low-energy models and metal-insulator transition}},
  author = {Cassol, Francesco and Gaspard, L\'eo and Casula, Michele and Martins, Cyril and Lenz, Benjamin},
  journal = {Phys. Rev. B},
  volume = {109},
  issue = {15},
  pages = {155120},
  numpages = {17},
  year = {2024},
  month = {Apr},
  publisher = {American Physical Society},
  doi = {10.1103/PhysRevB.109.155120},
  url = {https://link.aps.org/doi/10.1103/PhysRevB.109.155120}
}

@article{Kim_2008,
  title = {{Novel ${J}_{\mathrm{eff}}=1/2$ Mott State Induced by Relativistic Spin-Orbit Coupling in $\text{Sr}_2 \text{IrO}_4$}},
  author = {Kim, B. J. and Jin, Hosub and Moon, S. J. and Kim, J.-Y. and Park, B.-G. and Leem, C. S. and Yu, Jaejun and Noh, T. W. and Kim, C. and Oh, S.-J. and Park, J.-H. and Durairaj, V. and Cao, G. and Rotenberg, E.},
  journal = {Phys. Rev. Lett.},
  volume = {101},
  issue = {7},
  pages = {076402},
  numpages = {4},
  year = {2008},
  month = {Aug},
  publisher = {American Physical Society},
  url = {https://link.aps.org/doi/10.1103/PhysRevLett.101.076402}
}

@article{Kim_2009,
  title = {{Phase-sensitive observation of a
spin–orbital Mott state in $\text{Sr}_2 \text{IrO}_4$}},
  author = {Kim, B. J. and Ohsumi, H. and Komesu, T. and Sakai, S. and Morita, T. and Takagi, H.
and Arima, T. },
  journal = {Science},
  volume = {323},
  pages = {1239},
  year = {2009},
  month = {Mar},
  doi = {10.1126/science.1167106},
}

@article{Jeaong_2020,
  title = {{Magnetization Density Distribution of $\text{Sr}_2 \text{IrO}_4$: Deviation from a Local ${j}_{\mathrm{eff}}=1/2$ Picture}},
  author = {Jeong, Jaehong and Lenz, Benjamin and Gukasov, Arsen and Fabr\`eges, Xavier and Sazonov, Andrew and Hutanu, Vladimir and Louat, Alex and Bounoua, Dalila and Martins, Cyril and Biermann, Silke and Brouet, V\'eronique and Sidis, Yvan and Bourges, Philippe},
  journal = {Phys. Rev. Lett.},
  volume = {125},
  issue = {9},
  pages = {097202},
  numpages = {6},
  year = {2020},
  month = {Aug},
  publisher = {American Physical Society},
  doi = {10.1103/PhysRevLett.125.097202},
  url = {https://link.aps.org/doi/10.1103/PhysRevLett.125.097202}
}

@article{Uchida_2014,
  title = {{Correlated vs. conventional insulating behavior in the ${\mathrm{J}}_{\mathrm{eff}}=\frac{1}{2}$ vs. $\frac{3}{2}$ bands in the layered iridate $\text{Ba}_2 \text{IrO}_4$}},
  author = {Uchida, M. and Nie, Y. F. and King, P. D. C. and Kim, C. H. and Fennie, C. J. and Schlom, D. G. and Shen, K. M.},
  journal = {Phys. Rev. B},
  volume = {90},
  issue = {7},
  pages = {075142},
  numpages = {6},
  year = {2014},
  month = {Aug},
  publisher = {American Physical Society},
  url = {https://link.aps.org/doi/10.1103/PhysRevB.90.075142}
}

@article{Nichols_2014,
  title = {{Epitaxial $\text{Ba}_2 \text{IrO}_4$ thin-films grown on SrTiO 3 substrates by pulsed laser deposition}},
  author = {J. Nichols, J and Korneta, O. B. and  Terzic, J.  and  Cao , G. and Brill, J. W. and Seo, S. S. A.},
  journal = {Appl. Phys. Lett.},
  pages={121913},
  volume = {104},
  year = {2014},
  month = {March},
  url = {https://pubs.aip.org/aip/apl/article-abstract/104/12/121913/24554/Epitaxial-Ba2IrO4-thin-films-grown-on-SrTiO3?redirectedFrom=fulltext},
}

@article{Okabe_2012,
url = {https://dx.doi.org/10.1088/1742-6596/400/3/032071},
year = {2012},
month = {dec},
publisher = {},
volume = {400},
number = {3},
pages = {032071},
author = {H Okabe and M Isobe and E Takayama-Muromachi and A Koda and S Takeshita and M Hiraishi and M Miyazaki and R Kadono and Y Miyake and J Akimitsu},
title = {{Magnetic ordering in spin-orbit Mott insulator $\text{Ba}_2 \text{IrO}_4$ probed by $\mu$SR}},
journal = {Journal of Physics: Conference Series}
}

@article{DelaTorre_2015,
  title = {{Collapse of the Mott Gap and Emergence of a Nodal Liquid in Lightly Doped $\text{Ba}_2 \text{IrO}_4$}},
  author = {de la Torre, A. and McKeown Walker, S. and Bruno, F. Y. and Ricc\'o, S. and Wang, Z. and Gutierrez Lezama, I. and Scheerer, G. and Giriat, G. and Jaccard, D. and Berthod, C. and Kim, T. K. and Hoesch, M. and Hunter, E. C. and Perry, R. S. and Tamai, A. and Baumberger, F.},
  journal = {Phys. Rev. Lett.},
  volume = {115},
  issue = {17},
  pages = {176402},
  numpages = {5},
  year = {2015},
  month = {Oct},
  publisher = {American Physical Society},
  doi = {10.1103/PhysRevLett.115.176402},
  url = {https://link.aps.org/doi/10.1103/PhysRevLett.115.176402}
}

@article{CTQMC,
  title = {{Continuous-time Monte Carlo methods for quantum impurity models}},
  author = {Gull, Emanuel and Millis, Andrew J. and Lichtenstein, Alexander I. and Rubtsov, Alexey N. and Troyer, Matthias and Werner, Philipp},
  journal = {Rev. Mod. Phys.},
  volume = {83},
  issue = {2},
  pages = {349--404},
  numpages = {0},
  year = {2011},
  month = {May},
  publisher = {American Physical Society},
  doi = {10.1103/RevModPhys.83.349},
  url = {https://link.aps.org/doi/10.1103/RevModPhys.83.349}
}

@article{QuantumEspresso1,
  title = {{{QUANTUM ESPRESSO}}: A Modular and Open-Source Software Project for Quantum Simulations of Materials},
  shorttitle = {{{QUANTUM ESPRESSO}}},
  author = {Giannozzi, Paolo and Baroni, Stefano and Bonini, Nicola and Calandra, Matteo and Car, Roberto and Cavazzoni, Carlo and Ceresoli, Davide and Chiarotti, Guido L. and Cococcioni, Matteo and Dabo, Ismaila and Dal Corso, Andrea and {de Gironcoli}, Stefano and Fabris, Stefano and Fratesi, Guido and Gebauer, Ralph and Gerstmann, Uwe and Gougoussis, Christos and Kokalj, Anton and Lazzeri, Michele and {Martin-Samos}, Layla and Marzari, Nicola and Mauri, Francesco and Mazzarello, Riccardo and Paolini, Stefano and Pasquarello, Alfredo and Paulatto, Lorenzo and Sbraccia, Carlo and Scandolo, Sandro and Sclauzero, Gabriele and Seitsonen, Ari P. and Smogunov, Alexander and Umari, Paolo and Wentzcovitch, Renata M.},
  year = {2009},
  month = sep,
  journal = {Journal of Physics: Condensed Matter},
  volume = {21},
  number = {39},
  pages = {395502},
  issn = {1361-648X},
  doi = {https://doi.org/10.1088/0953-8984/21/39/395502},
  langid = {english},
  pmid = {21832390},
}

@article{TRIQS2015,
  title = {{{{TRIQS}}: {{A}} Toolbox for Research on Interacting Quantum Systems}},
  shorttitle = {{{TRIQS}}},
  author = {Parcollet, Olivier and Ferrero, Michel and Ayral, Thomas and Hafermann, Hartmut and Krivenko, Igor and Messio, Laura and Seth, Priyanka},
  year = {2015},
  month = nov,
  journal = {Computer Physics Communications},
  volume = {196},
  pages = {398--415},
  issn = {0010-4655},
  doi = {https://doi.org/10.1016/j.cpc.2015.04.023},
  urldate = {2023-05-09},
  langid = {english}
}

@article{Arita_2012,
  title = {{Ab initio Studies on the Interplay between Spin-Orbit Interaction and Coulomb Correlation in $\text{Sr}_2 \text{IrO}$ and $\text{Ba}_2 \text{IrO}_4$}},
  author = {Arita, R. and Kune\ifmmode \check{s}\else \v{s}\fi{}, J. and Kozhevnikov, A. V. and Eguiluz, A. G. and Imada, M.},
  journal = {Phys. Rev. Lett.},
  volume = {108},
  issue = {8},
  pages = {086403},
  numpages = {5},
  year = {2012},
  month = {Feb},
  publisher = {American Physical Society},
  doi = {10.1103/PhysRevLett.108.086403},
  url = {https://link.aps.org/doi/10.1103/PhysRevLett.108.086403}
}

@article{Damascelli_2021,
  title = {Constraints on the two-dimensional pseudospin-$\frac{1}{2}$ Mott insulator description of ${\mathrm{Sr}}_{2}{\mathrm{IrO}}_{4}$},
  author = {Zwartsenberg, B. and Day, R. P. and Razzoli, E. and Michiardi, M. and Na, M. X. and Zhang, G. and Denlinger, J. D. and Vobornik, I. and Bigi, C. and Kim, B. J. and Elfimov, I. S. and Pavarini, E. and Damascelli, A.},
  journal = {Phys. Rev. B},
  volume = {105},
  issue = {24},
  pages = {245130},
  numpages = {9},
  year = {2022},
  month = {Jun},
  publisher = {American Physical Society},
  doi = {10.1103/PhysRevB.105.245130},
  url = {https://link.aps.org/doi/10.1103/PhysRevB.105.245130}
}

@article{Pavarini_2023,
  title = {{Multiorbital Nature of Doped $\text{Sr}_2 \text{IrO}_4$}},
  author = {Zhang, Guoren and Pavarini, Eva},
  journal = {Phys. Rev. Lett.},
  volume = {131},
  issue = {3},
  pages = {036504},
  numpages = {6},
  year = {2023},
  month = {Jul},
  publisher = {American Physical Society},
  doi = {10.1103/PhysRevLett.131.036504},
  url = {https://link.aps.org/doi/10.1103/PhysRevLett.131.036504}
}

@article{Pavarini_2025,
  title = {{Nature of the high-pressure insulating state in ${\mathrm{Sr}}_{2}{\mathrm{IrO}}_{4}$: Mott picture}},
  author = {Zhang, Guoren and Hadipour, Hanif and Pavarini, Eva},
  journal = {Phys. Rev. B},
  volume = {111},
  issue = {12},
  pages = {L121101},
  numpages = {6},
  year = {2025},
  month = {Mar},
  publisher = {American Physical Society},
  doi = {10.1103/PhysRevB.111.L121101},
  url = {https://link.aps.org/doi/10.1103/PhysRevB.111.L121101}
}

@article{Pavarini_2021,
  title = {{Magnetic superexchange couplings in ${\mathrm{Sr}}_{2}{\mathrm{IrO}}_{4}$}},
  author = {Zhang, Guoren and Pavarini, Eva},
  journal = {Phys. Rev. B},
  volume = {104},
  issue = {12},
  pages = {125116},
  numpages = {13},
  year = {2021},
  month = {Sep},
  publisher = {American Physical Society},
  doi = {10.1103/PhysRevB.104.125116},
  url = {https://link.aps.org/doi/10.1103/PhysRevB.104.125116}
}

@article{Lenz_2019,
doi = {10.1088/1361-648X/ab146a},
url = {https://dx.doi.org/10.1088/1361-648X/ab146a},
year = {2019},
month = {may},
publisher = {IOP Publishing},
volume = {31},
number = {29},
pages = {293001},
author = {Lenz, B and Martins, C and Biermann, S},
title = {{Spectral functions of $\text{Sr}_2 \text{IrO}_4$: theory versus experiment}},
journal = {Journal of Physics: Condensed Matter}
}

@article{Isobe_2012,
doi = {10.1088/1742-6596/400/3/032028},
url = {https://dx.doi.org/10.1088/1742-6596/400/3/032028},
year = {2012},
month = {dec},
publisher = {},
volume = {400},
number = {3},
pages = {032028},
author = {Isobe, M and Okabe, H and Takayama-Muromachi, E and Koda, A and Takeshita, S and Hiraishi, M and Miyazaki, M and Kadono, R and Miyake, Y and Akimitsu, J},
title = {{Spin-Orbit Mott State in the Novel Quasi-2D Antiferromagnet $\text{Ba}_2 \text{IrO}_4$}},
journal = {Journal of Physics: Conference Series}
}

@article{Liu_2018,
  title = {Relativistic $GW$+BSE study of the optical properties of Ruddlesden-Popper iridates},
  author = {Liu, Peitao and Kim, Bongjae and Chen, Xing-Qiu and Sarma, D. D. and Kresse, Georg and Franchini, Cesare},
  journal = {Phys. Rev. Mater.},
  volume = {2},
  issue = {7},
  pages = {075003},
  numpages = {13},
  year = {2018},
  month = {Jul},
  publisher = {American Physical Society},
  doi = {10.1103/PhysRevMaterials.2.075003},
  url = {https://link.aps.org/doi/10.1103/PhysRevMaterials.2.075003}
}

@article{Zhang_2013,
  title = {{Effective $J\mathbf{=}1/2$ Insulating State in Ruddlesden-Popper Iridates: An $\mathrm{LDA}\mathbf{+}\mathrm{DMFT}$ Study}},
  author = {Zhang, Hongbin and Haule, Kristjan and Vanderbilt, David},
  journal = {Phys. Rev. Lett.},
  volume = {111},
  issue = {24},
  pages = {246402},
  numpages = {5},
  year = {2013},
  month = {Dec},
  publisher = {American Physical Society},
  doi = {10.1103/PhysRevLett.111.246402},
  url = {https://link.aps.org/doi/10.1103/PhysRevLett.111.246402}
}

@article{Moser_2014, 
  title = {{The electronic structure of the high-symmetry perovskite iridate $\text{Ba}_2\text{IrO}_4$}},
  author = {Moser, S. and Moreschini, L. and Ebrahimi, A. and Dalla Piazza,  B.  and Isobe, M. and Okabe, H. and Akimitsu, J. and Mazurenko, V. V. and Kim, K. S. and Bostwick, A. and Rotenberg, E. and Chang, J. and R{\o}nnow, H. M. and Grioni, M.},
  year = {2014},
  month = feb,
  journal = {New Journal of Physics},
  volume = {16},
  pages = {013008},
  url={https://iopscience.iop.org/article/10.1088/1367-2630/16/1/013008},
  urldate = {2014-09-01},
  langid = {english},
}

@article{Dai_2014,
  title = {{Local density of states study of a spin-orbit-coupling induced Mott insulator ${\mathrm{Sr}}_{2}\mathrm{Ir}{\mathrm{O}}_{4}$}},
  author = {Dai, Jixia and Calleja, Eduardo and Cao, Gang and McElroy, Kyle},
  journal = {Phys. Rev. B},
  volume = {90},
  issue = {4},
  pages = {041102},
  numpages = {5},
  year = {2014},
  month = {Jul},
  publisher = {American Physical Society},
  doi = {10.1103/PhysRevB.90.041102},
  url = {https://link.aps.org/doi/10.1103/PhysRevB.90.041102}
}

@article{Brouet_2015,
  title = {{Transfer of spectral weight across the gap of ${\mathrm{Sr}}_{2}{\mathrm{IrO}}_{4}$ induced by La doping}},
  author = {Brouet, V\'eronique and Mansart, Joseph and Perfetti, Luca and Piovera, Christian and Vobornik, Ivana and Le F\`evre, Patrick and Bertran, Fran\c{c}ois and Riggs, Scott C. and Shapiro, M. C. and Giraldo-Gallo, Paula and Fisher, Ian R.},
  journal = {Phys. Rev. B},
  volume = {92},
  issue = {8},
  pages = {081117},
  numpages = {5},
  year = {2015},
  month = {Aug},
  publisher = {American Physical Society},
  doi = {10.1103/PhysRevB.92.081117},
  url = {https://link.aps.org/doi/10.1103/PhysRevB.92.081117}
}

@article{Georges_1996,
  title = {{Dynamical mean-field theory of strongly correlated fermion systems and the limit of infinite dimensions}},
  author = {Georges, Antoine and Kotliar, Gabriel and Krauth, Werner and Rozenberg, Marcelo J.},
  journal = {Rev. Mod. Phys.},
  volume = {68},
  issue = {1},
  pages = {13--125},
  numpages = {0},
  year = {1996},
  month = {Jan},
  publisher = {American Physical Society},
  doi = {10.1103/RevModPhys.68.13},
  url = {https://link.aps.org/doi/10.1103/RevModPhys.68.13}
}

@article{Sangiovanni_2006,
  title = {{Static versus dynamical mean-field theory of Mott antiferromagnets}},
  author = {Sangiovanni, G. and Toschi, A. and Koch, E. and Held, K. and Capone, M. and Castellani, C. and Gunnarsson, O. and Mo, S.-K. and Allen, J. W. and Kim, H.-D. and Sekiyama, A. and Yamasaki, A. and Suga, S. and Metcalf, P.},
  journal = {Phys. Rev. B},
  volume = {73},
  issue = {20},
  pages = {205121},
  numpages = {5},
  year = {2006},
  month = {May},
  publisher = {American Physical Society},
  doi = {10.1103/PhysRevB.73.205121},
  url = {https://link.aps.org/doi/10.1103/PhysRevB.73.205121}
}

@article{
Choi_2024,
author = {Dongsung Choi  and Changming Yue  and Doron Azoury  and Zachary Porter  and Jiyu Chen  and Francesco Petocchi  and Edoardo Baldini  and Baiqing Lv  and Masataka Mogi  and Yifan Su  and Stephen D. Wilson  and Martin Eckstein  and Philipp Werner  and Nuh Gedik },
title = {{Light-induced insulator-metal transition in Sr$_2$IrO$_4$ reveals the nature of the insulating ground state}},
journal = {Proceedings of the National Academy of Sciences},
volume = {121},
number = {29},
pages = {e2323013121},
year = {2024},
doi = {10.1073/pnas.2323013121},
}

@article{Foulquier_2023,
  author = {Foulquier, Paul and Civelli, Marcello and Rozenberg, Marcelo and Camjayi, Alberto and Bobadilla, Joel and Colson, Dorothée and Forget, Anne and Thuéry, Pierre and Bertran, François and Le Fèvre, Patrick and Brouet, Véronique},
  title = {Evolution of the spectral lineshape at the magnetic transition in Sr$_2$IrO$_4$ and Sr$_3$Ir$_2$O$_7$},
  journal = {The European Physical Journal B},
  year = {2023},
  volume = {96},
  number = {4},
  pages = {42},
  doi = {10.1140/epjb/s10051-023-00512-3},
  url = {https://doi.org/10.1140/epjb/s10051-023-00512-3}
}

@article{Fratino_2017,
  title = {Signatures of the Mott transition in the antiferromagnetic state of the two-dimensional Hubbard model},
  author = {Fratino, L. and S\'emon, P. and Charlebois, M. and Sordi, G. and Tremblay, A.-M. S.},
  journal = {Phys. Rev. B},
  volume = {95},
  issue = {23},
  pages = {235109},
  numpages = {11},
  year = {2017},
  month = {Jun},
  publisher = {American Physical Society},
  doi = {10.1103/PhysRevB.95.235109},
  url = {https://link.aps.org/doi/10.1103/PhysRevB.95.235109}
}

@article{Li_2013,
  author    = {Li, Qing and Cao, Guixin and Okamoto, Satoshi and Yi, Jieyu and Lin, Wenzhi and Sales, Brian C. and Yan, Jiaqiang and Arita, Ryotaro and Kune{\v{s}}, Jan and Kozhevnikov, Anton V. and Eguiluz, Adolfo G. and Imada, Masatoshi and Gai, Zheng and Pan, Minghu and Mandrus, David G.},
  title     = {Atomically resolved spectroscopic study of Sr2IrO4: Experiment and theory},
  journal   = {Scientific Reports},
  year      = {2013},
  volume    = {3},
  number    = {1},
  pages     = {3073},
  doi       = {10.1038/srep03073},
  url       = {https://doi.org/10.1038/srep03073}
}

@article{Cassol_2026,
  title = {Spin-polaron fingerprints in the optical conductivity of iridates},
  author = {Cassol, Francesco and Gaspard, L\'eo and Martins, Cyril and Casula, Michele and Lenz, Benjamin},
  journal = {Phys. Rev. B},
  volume = {113},
  issue = {24},
  pages = {L241107},
  numpages = {8},
  year = {2026},
  month = {Jun},
  publisher = {American Physical Society},
  doi = {10.1103/xt5x-8c92},
  url = {https://link.aps.org/doi/10.1103/xt5x-8c92}
}

@article{MEM_1,
  title = {Quantum Monte Carlo simulations and maximum entropy: Dynamics from imaginary-time data},
  author = {Gubernatis, J. E. and Jarrell, Mark and Silver, R. N. and Sivia, D. S.},
  journal = {Phys. Rev. B},
  volume = {44},
  issue = {12},
  pages = {6011--6029},
  numpages = {0},
  year = {1991},
  month = {Sep},
  publisher = {American Physical Society},
  doi = {10.1103/PhysRevB.44.6011},
  url = {https://link.aps.org/doi/10.1103/PhysRevB.44.6011}
}

@article{Kim_2020,
  title = {Alleviating the sign problem in quantum Monte Carlo simulations of spin-orbit-coupled multiorbital Hubbard models},
  author = {Kim, Aaram J. and Werner, Philipp and Valent\'{\i}, Roser},
  journal = {Phys. Rev. B},
  volume = {101},
  issue = {4},
  pages = {045108},
  numpages = {10},
  year = {2020},
  month = {Jan},
  publisher = {American Physical Society},
  doi = {10.1103/PhysRevB.101.045108},
  url = {https://link.aps.org/doi/10.1103/PhysRevB.101.045108}
}

@article{Chatzieleftheriou_2024,
  title = {Local and Nonlocal Electronic Correlations at the Metal-Insulator Transition in the Two-Dimensional Hubbard Model},
  author = {Chatzieleftheriou, Maria and Biermann, Silke and Stepanov, Evgeny A.},
  journal = {Phys. Rev. Lett.},
  volume = {132},
  issue = {23},
  pages = {236504},
  numpages = {7},
  year = {2024},
  month = {Jun},
  publisher = {American Physical Society},
  doi = {10.1103/PhysRevLett.132.236504},
  url = {https://link.aps.org/doi/10.1103/PhysRevLett.132.236504}
}

@article{Staudt_2000,
  author    = {R. Staudt and M. Dzierzawa and A. Muramatsu},
  title     = {Phase diagram of the three-dimensional Hubbard model at half filling},
  journal   = {The European Physical Journal B},
  year      = {2000},
  volume    = {17},
  number    = {3},
  pages     = {411--415},
  doi       = {10.1007/s100510070120},
  url       = {https://doi.org/10.1007/s100510070120},
  issn      = {1434-6036}
}

@article{Rohringer_2011,
  title = {Critical Properties of the Half-Filled Hubbard Model in Three Dimensions},
  author = {Rohringer, G. and Toschi, A. and Katanin, A. and Held, K.},
  journal = {Phys. Rev. Lett.},
  volume = {107},
  issue = {25},
  pages = {256402},
  numpages = {5},
  year = {2011},
  month = {Dec},
  publisher = {American Physical Society},
  doi = {10.1103/PhysRevLett.107.256402},
  url = {https://link.aps.org/doi/10.1103/PhysRevLett.107.256402}
}

@article{Jarrell_1992,
  title = {Hubbard model in infinite dimensions: A quantum Monte Carlo study},
  author = {Jarrell, M.},
  journal = {Phys. Rev. Lett.},
  volume = {69},
  issue = {1},
  pages = {168--171},
  numpages = {0},
  year = {1992},
  month = {Jul},
  publisher = {American Physical Society},
  doi = {10.1103/PhysRevLett.69.168},
  url = {https://link.aps.org/doi/10.1103/PhysRevLett.69.168}
}

@article{Vucevic_2013,
  title = {Finite-temperature crossover and the quantum Widom line near the Mott transition},
  author = {Vu\ifmmode \check{c}\else \v{c}\fi{}i\ifmmode \check{c}\else \v{c}\fi{}evi\ifmmode \acute{c}\else \'{c}\fi{}, J. and Terletska, H. and Tanaskovi\ifmmode \acute{c}\else \'{c}\fi{}, D. and Dobrosavljevi\ifmmode \acute{c}\else \'{c}\fi{}, V.},
  journal = {Phys. Rev. B},
  volume = {88},
  issue = {7},
  pages = {075143},
  numpages = {12},
  year = {2013},
  month = {Aug},
  publisher = {American Physical Society},
  doi = {10.1103/PhysRevB.88.075143},
}

@article{Terletska_2011,
  title = {Quantum Critical Transport near the Mott Transition},
  author = {Terletska, H. and Vu\ifmmode \check{c}\else \v{c}\fi{}i\ifmmode \check{c}\else \v{c}\fi{}evi\ifmmode \acute{c}\else \'{c}\fi{}, J. and Tanaskovi\ifmmode \acute{c}\else \'{c}\fi{}, D. and Dobrosavljevi\ifmmode \acute{c}\else \'{c}\fi{}, V.},
  journal = {Phys. Rev. Lett.},
  volume = {107},
  issue = {2},
  pages = {026401},
  numpages = {4},
  year = {2011},
  month = {Jul},
  publisher = {American Physical Society},
  doi = {10.1103/PhysRevLett.107.026401},
  url = {https://link.aps.org/doi/10.1103/PhysRevLett.107.026401}
}

@article{Crispino_2023,
  title = {Slave-spin mean field for broken-symmetry states: N\'eel antiferromagnetism and its phase separation in multiorbital Hubbard models},
  author = {Crispino, Matteo and Chatzieleftheriou, Maria and Gorni, Tommaso and de' Medici, Luca},
  journal = {Phys. Rev. B},
  volume = {107},
  issue = {15},
  pages = {155149},
  numpages = {14},
  year = {2023},
  month = {Apr},
  publisher = {American Physical Society},
  doi = {10.1103/PhysRevB.107.155149},
  url = {https://link.aps.org/doi/10.1103/PhysRevB.107.155149}
}

@article{Taranto_2012,
  title = {Signature of antiferromagnetic long-range order in the optical spectrum of strongly correlated electron systems},
  author = {Taranto, C. and Sangiovanni, G. and Held, K. and Capone, M. and Georges, A. and Toschi, A.},
  journal = {Phys. Rev. B},
  volume = {85},
  issue = {8},
  pages = {085124},
  numpages = {10},
  year = {2012},
  month = {Feb},
  publisher = {American Physical Society},
  doi = {10.1103/PhysRevB.85.085124},
  url = {https://link.aps.org/doi/10.1103/PhysRevB.85.085124},
}

@article{Hsieh_2012,
  title = {Observation of a metal-to-insulator transition with both Mott-Hubbard and Slater characteristics in Sr${}_{2}$IrO${}_{4}$ from time-resolved photocarrier dynamics},
  author = {Hsieh, D. and Mahmood, F. and Torchinsky, D. H. and Cao, G. and Gedik, N.},
  journal = {Phys. Rev. B},
  volume = {86},
  issue = {3},
  pages = {035128},
  numpages = {6},
  year = {2012},
  month = {Jul},
  publisher = {American Physical Society},
  doi = {10.1103/PhysRevB.86.035128},
  url = {https://link.aps.org/doi/10.1103/PhysRevB.86.035128},
}

@article{Lifshitz,
  title = {Anomalies of Electron Characteristics of a Metal in the High Pressure Region.},
  author = {Lifshitz, I.M.},
  journal = {Soviet Physics JETP},
  volume = {11},
  pages={1130},
  year = {1960},
  url = {http://www.jetp.ras.ru/cgi-bin/dn/e_011_05_1130.pdf}
}

@article{Zhang_2016,
  title = {Fermi Surface of ${\mathrm{Sr}}_{2}{\mathrm{RuO}}_{4}$: Spin-Orbit and Anisotropic Coulomb Interaction Effects},
  author = {Zhang, Guoren and Gorelov, Evgeny and Sarvestani, Esmaeel and Pavarini, Eva},
  journal = {Phys. Rev. Lett.},
  volume = {116},
  issue = {10},
  pages = {106402},
  numpages = {6},
  year = {2016},
  month = {Mar},
  publisher = {American Physical Society},
  doi = {10.1103/PhysRevLett.116.106402},
  url = {https://link.aps.org/doi/10.1103/PhysRevLett.116.106402}
}

@misc{Vucicevic_2026,
    title = {{Hubbard} vs. {Emery} model: spectra, transport and relevance for cuprates},
    author={Jakša Vučičević and Rok Žitko},
    year = {2026},
    eprint = {2604.08085},
    archivePrefix = {arXiv},
    primaryClass = {cond-mat.str-el},
    url = {https://arxiv.org/abs/2604.08085},
}

@article{Martins_2017,
doi = {10.1088/1361-648X/aa648f},
url = {https://doi.org/10.1088/1361-648X/aa648f},
year = {2017},
month = {may},
publisher = {IOP Publishing},
volume = {29},
number = {26},
pages = {263001},
author = {Martins, C and Aichhorn, M and Biermann, S},
title = {Coulomb correlations in 4d and 5d oxides from first principles—or how spin–orbit materials choose their effective orbital degeneracies},
journal = {Journal of Physics: Condensed Matter}
}

@article{Slater_1951,
  author  = {Slater, J. C.},
  title   = {Magnetic Effects and the {H}artree-{F}ock Equation},
  journal = {Phys. Rev.},
  volume  = {82},
  pages   = {538--541},
  year    = {1951},
  doi     = {10.1103/PhysRev.82.538}
}

@article{Pruschke_2003,
  author  = {Pruschke, Th. and Zitzler, R.},
  title   = {From {S}later to {M}ott-{H}eisenberg physics: the antiferromagnetic phase of the {H}ubbard model},
  journal = {J. Phys.: Condens. Matter},
  volume  = {15},
  pages   = {7867},
  year    = {2003}
}

@article{Watanabe_2014,
  author  = {Watanabe, Hiroshi and Shirakawa, Tomonori and Yunoki, Seiji},
  title   = {Theoretical study of insulating mechanism in multiorbital {H}ubbard models with a large spin-orbit coupling: {S}later versus {M}ott scenario in {Sr$_2$IrO$_4$}},
  journal = {Phys. Rev. B},
  volume  = {89},
  pages   = {165115},
  year    = {2014},
  doi     = {10.1103/PhysRevB.89.165115}
}

@article{Chubukov_1992,
  author  = {Chubukov, Andrey V. and Frenkel, David M.},
  title   = {Renormalized perturbation theory of magnetic instabilities in the two-dimensional {H}ubbard model at small doping},
  journal = {Phys. Rev. B},
  volume  = {46},
  pages   = {11884--11901},
  year    = {1992},
  doi     = {10.1103/PhysRevB.46.11884}
}

@article{JKim_2012,
  author  = {Kim, Jungho and Casa, D. and Upton, M. H. and Gog, T. and Kim, Young-June and Mitchell, J. F. and van Veenendaal, M. and Daghofer, M. and van den Brink, J. and Khaliullin, G. and Kim, B. J.},
  title   = {Magnetic Excitation Spectra of {Sr$_2$IrO$_4$} Probed by Resonant Inelastic X-Ray Scattering: Establishing Links to Cuprate Superconductors},
  journal = {Phys. Rev. Lett.},
  volume  = {108},
  pages   = {177003},
  year    = {2012},
  doi     = {10.1103/PhysRevLett.108.177003}
}

@article{Fujiyama_2012,
  author  = {Fujiyama, S. and Ohsumi, H. and Komesu, T. and Matsuno, J. and Kim, B. J. and Takata, M. and Arima, T. and Takagi, H.},
  title   = {Two-Dimensional {H}eisenberg Behavior of ${J}_{\mathrm{eff}}=1/2$ Isospins in the Paramagnetic State of the Spin-Orbital {M}ott Insulator {Sr$_2$IrO$_4$}},
  journal = {Phys. Rev. Lett.},
  volume  = {108},
  pages   = {247212},
  year    = {2012},
  doi     = {10.1103/PhysRevLett.108.247212}
}

@article{Crispino_2026,
  title = {Local-moment magnetism in Mn-based pnictides},
  author = {Crispino, Matteo and Witt, Niklas and Gorni, Tommaso and Sangiovanni, Giorgio and de' Medici, Luca},
  journal = {Phys. Rev. B},
  volume = {113},
  issue = {13},
  pages = {134434},
  numpages = {10},
  year = {2026},
  month = {Apr},
  publisher = {American Physical Society},
  doi = {10.1103/yzkk-71cr},
  url = {https://link.aps.org/doi/10.1103/yzkk-71cr}
}

@inbook{Seki_2013,
    title = {Temperature Dependence of the Optical Conductivity in a Half-Filled {Hubbard} Model: Mott-Type Insulator Vs. {Slater}-Type Insulator},
    author = {Seki, K. and Shirakawa, T. and Sun, Y. and Yunoki, S.},
    booktitle = {Proceedings of the International Conference on Strongly Correlated Electron Systems (SCES2013)},
    year = {2014},
    volume = {3},
    pages = {014026},
    numpages = {6},
    publisher = {The Physical Society of Japan},
    doi = {10.7566/JPSCP.3.014026},
}

@article{Raczkowski_2021,
  title = {Local moments versus itinerant antiferromagnetism: Magnetic phase diagram and spectral properties of the anisotropic square lattice Hubbard model},
  author = {Raczkowski, Marcin and Assaad, Fakher F. and Imada, Masatoshi},
  journal = {Phys. Rev. B},
  volume = {103},
  issue = {12},
  pages = {125137},
  numpages = {18},
  year = {2021},
  month = {Mar},
  publisher = {American Physical Society},
  doi = {10.1103/PhysRevB.103.125137},
  url = {https://link.aps.org/doi/10.1103/PhysRevB.103.125137}
}

\setcounter{figure}{0}
\thispagestyle{empty}

\renewcommand{\figurename}{}
\renewcommand{\tablename}{}
\renewcommand{\thefigure}{Extended Data Fig.~\arabic{figure}}
\renewcommand{\thetable}{Extended Data Tab.~\arabic{table}}

\clearpage

\begin{figure}[t!]
    \centering
    \includegraphics[width=0.88\linewidth]{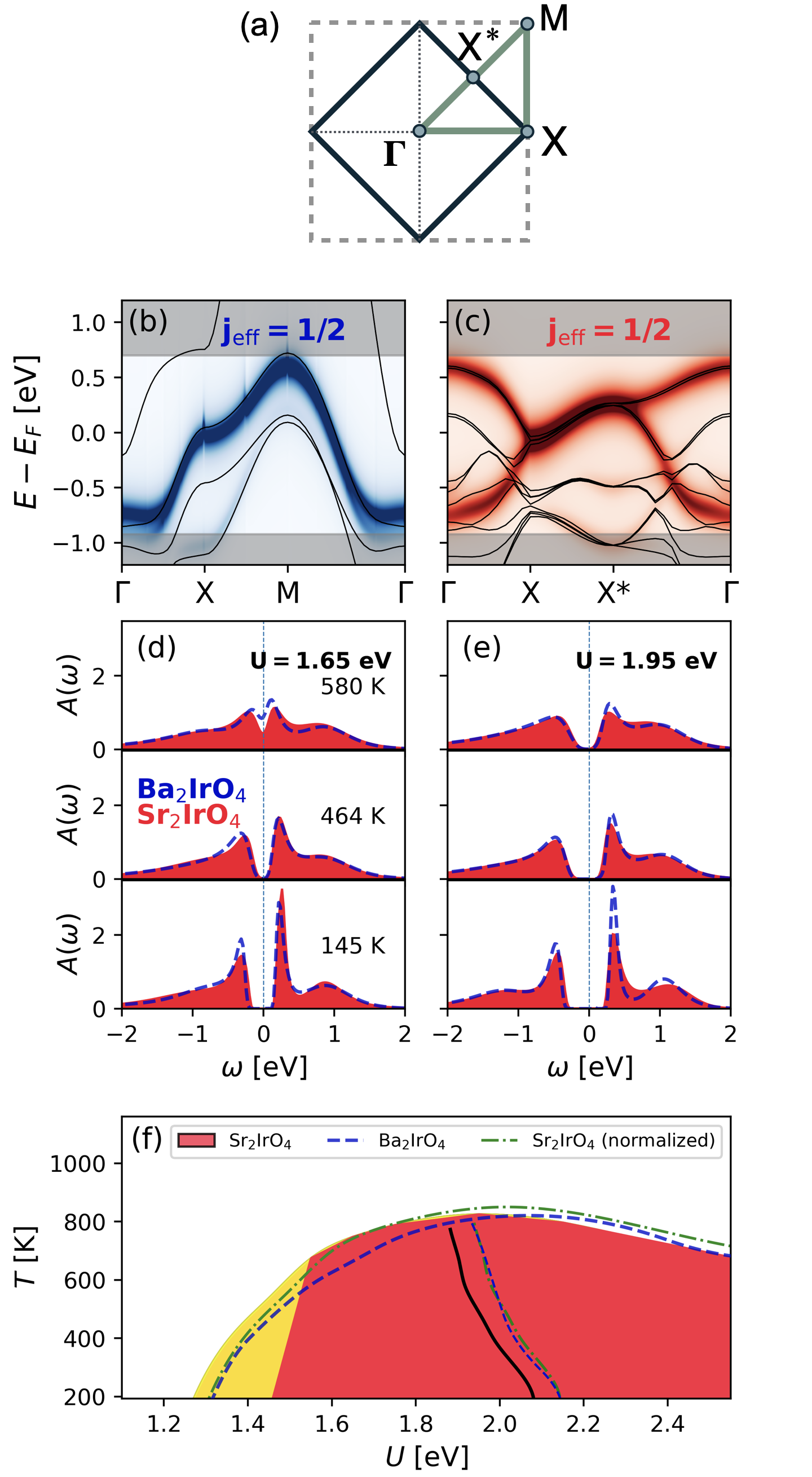}
    \caption{
   Comparison between \bairo\ and \sriro.
   (a) Brillouin zone of \bairo\ (large) and \sriro (small). 
   Dispersion of the \jeff\ band in our tight-binding model for \bairo (b) and \sriro (c). Notice that in \sriro\ structural distortions compensate the smaller unit cell, reducing the corresponding bandwidth $W$\cite{Martins_2011}. As a consequence, the extension of the \jeff\ band becomes comparable in both compounds.
   Panels (d) and (e) compare AFM-DMFT calculations of the two materials for different interaction strength and temperature, indicating similar behavior.
   The blue lines refer to \bairo, while red shaded areas correspond to \sriro.
   (f) Comparison between the \bairo\ and \sriro phase diagram. The filled red region denotes the \sriro\ AFM dome, with the corresponding Widom line reported in black ($U_w(S)$). 
   The yellow area indicates the coexistence region.
   Dashed blue lines identify $T_N$ and $U_W$ for \bairo.
   Dashed green curves represent \sriro\ transition lines rescaled by a factor 1.03, and coincide with the ones of \bairo. 
   }
       \vspace{5pt}
    \hrule
    \label{fig:Appendix_1}
\end{figure}

\begin{figure}[t!]
    \centering
    \includegraphics[width=0.88\linewidth]{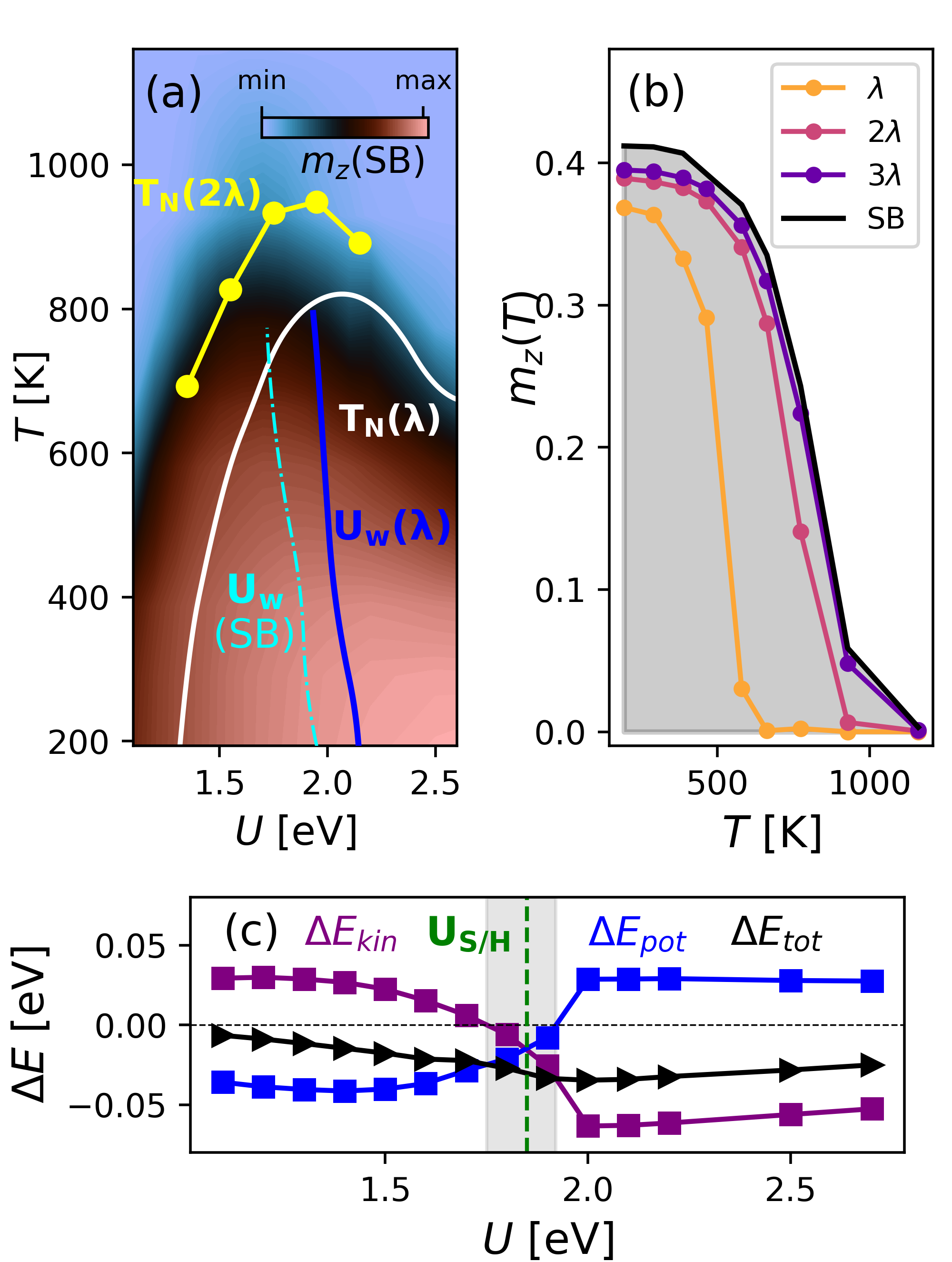}
    \caption{ Single-band \jeff\ and three-band $t_{2g}$ phase diagram comparison. 
   (a) Magnetization of the single band \jeff\ model in the $U-T$ space. White and yellow lines represent the transition line of the corresponding \textit{ab initio} $t_{2g}$ model and of the one obtained by doubling the SOC constant.  In (light)dark blue the Widom line for the (single-)three-band model is reported. The passage to the Mott state is found at lower $U$ in the single-band case, proving the active role of the \JEFF\ bands in screening the interactions.
   (b) Temperature profile of the magnetization $m_z(T)$ for the
   single-band model and for the three band model for different values
   of the SOC coupling $\lambda$, with $U = 1.55$ eV. (c) Energetics
   of the single band \jeff\ model. Also in this case, the passage to a localized type of magnetism occurs for $U<U_W$.
}
    \vspace{5pt}
    \hrule
    \label{fig:Appendix_2}
\end{figure}

\clearpage

\begin{figure}[t!]
    \centering
    \includegraphics[width=\linewidth]{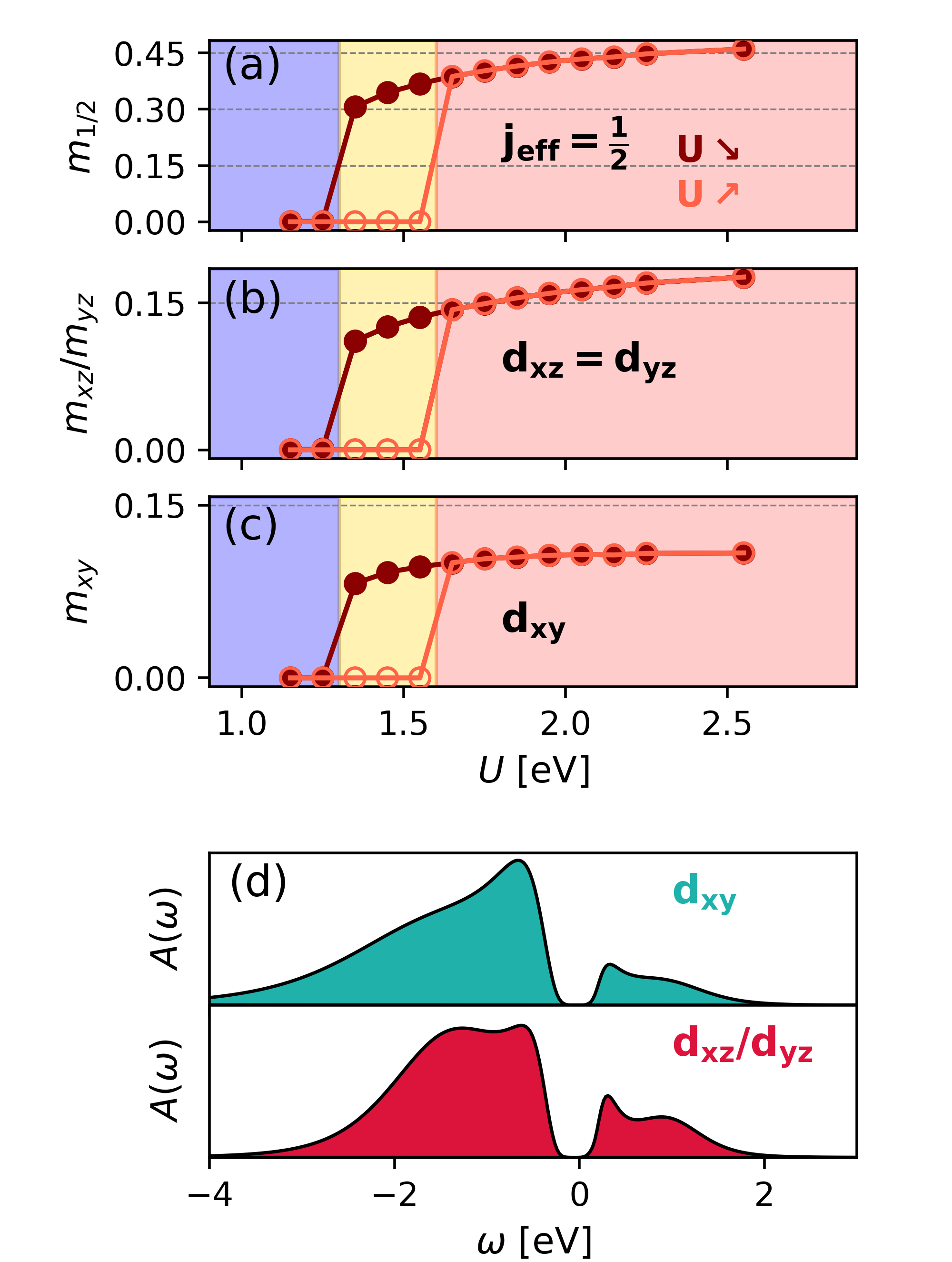}
    \caption{
     Observables from the $t_{2g}$ perspective. 
     Magnetization in the \jeff(a), $d_{xz}/d_{yz}$ (b) and $d_{xy}$ (c) channels respectively. (d) Spectral function projected on the three $t_{2g}$ orbitals.
    }
     \vspace{5pt}
    \hrule
    \label{fig:Appendix_3}
\end{figure}

\begin{figure}[t!]
    \centering
    \includegraphics[width=\linewidth]{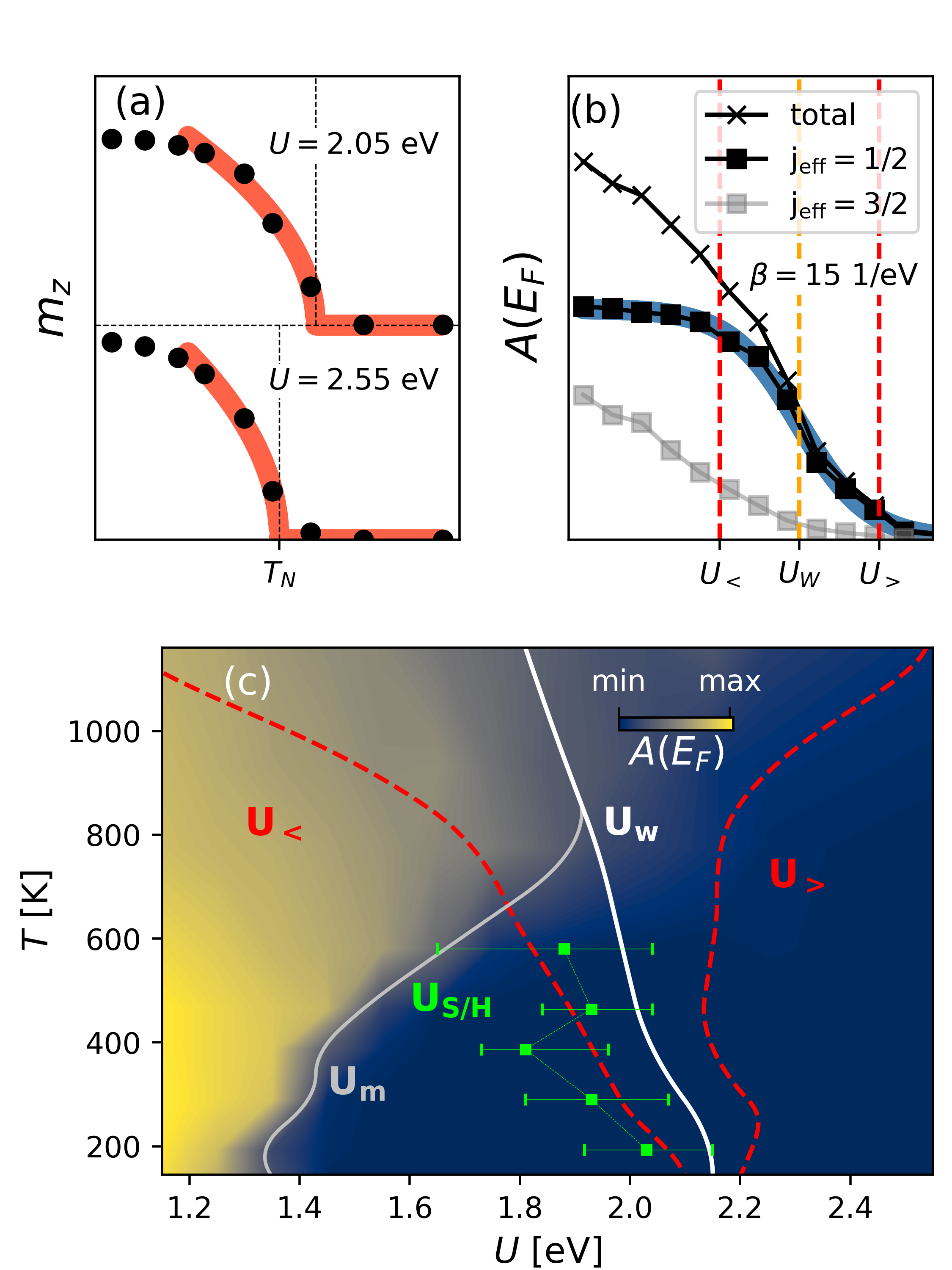}
    \caption{
Determination of the magnetic transition and the metal-insulator crossover region. 
    (a) Fit of the staggered magnetization $m_z$ for $U=2.05$ eV, and $2.55$ eV, respectively (red curve). 
    (b) Fit of the spectral function at $\beta= 15$ 1/eV (blue curve). 
    The orange dashed line indicates the middle of the crossover, while
    the crossover boundaries $U_{<}$ and $U_{>}$ are shown in red, 
    and are a guide to the eye.
    (c) Spectral function at the Fermi level $A(E_F)$. The $U_m$ and $U_W$ lines denote the passage from metallic to insulating phases for the AFM and PM solutions, respectively. As such, they merge above $T_N$, where magnetic solutions cannot be stabilized. The crossover region is reported by $U_<$ and $U_>$ for the enforced PM solution, and it narrows progressively upon cooling. 
    Its conical shape is the hallmark of the supercritical region above the critical endpoint of a first-order Mott transition~\cite{Terletska_2011,Vucevic_2013}, which in our model lies below the lowest temperature considered here, hidden beneath the AFM dome. The cooperation region is also reported in green, $U_{S/H}$ denoting its middle point. We notice that $U_{S/H}$ approximately coincides with $U_<$: when the AFM state becomes Heisenberg-driven, the corresponding PM solution starts displaying spectral weight suppression. 
    On the other hand, the end of the cooperation region, derived for $\Delta E_{pot}> 0$ eV, is consistent with the location of the Widom line.
}
    \vspace{5pt}
    \hrule
    \label{fig:Appendix_4}
\end{figure}

\clearpage
\newpage

\begin{figure}[t!]
    \centering
    \includegraphics[width=\linewidth]{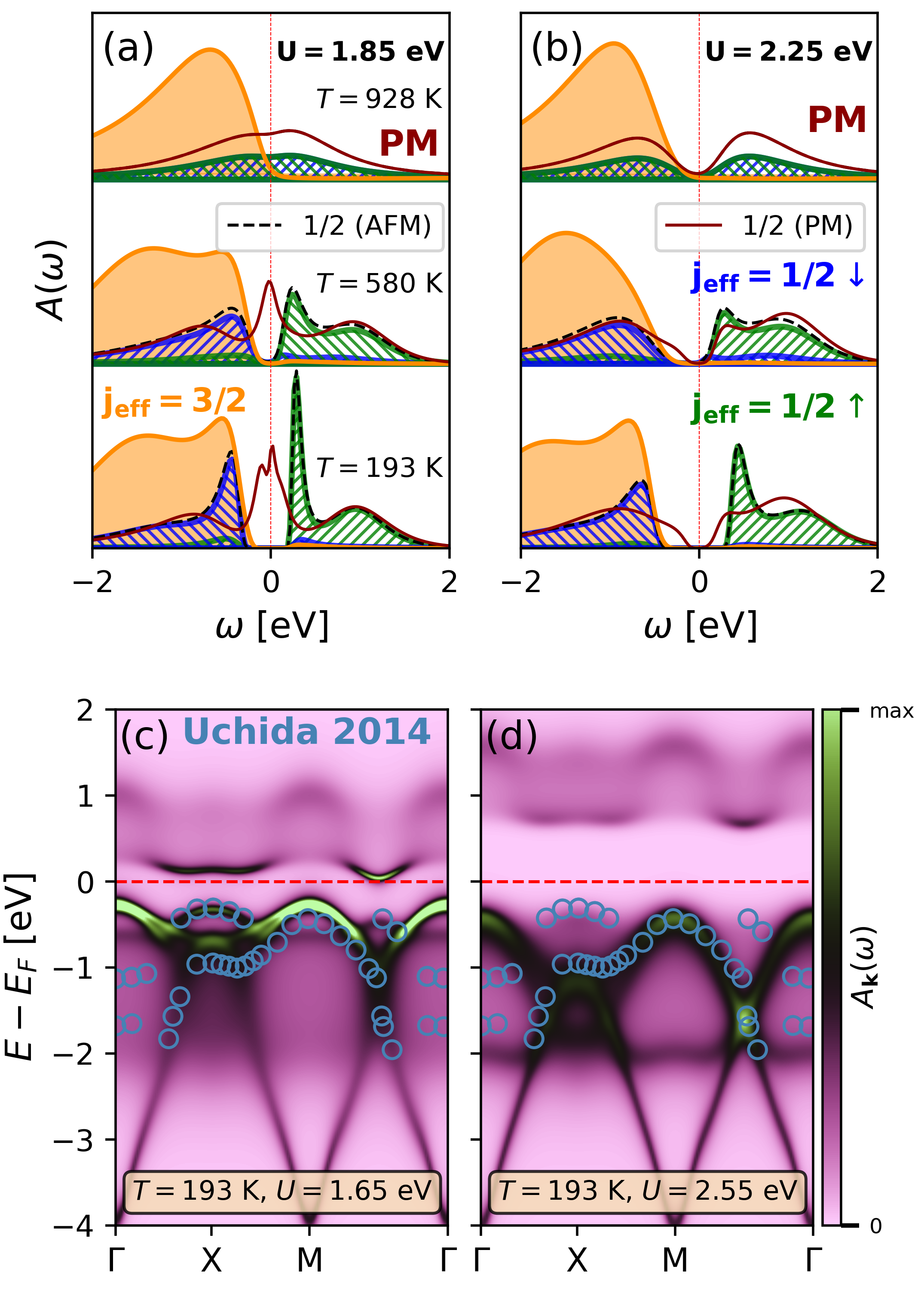}
    \caption{
Additional spectral features of iridates.  
Total density of states in proximity of the beginning (a) and of the end (b) of the cooperation region. For $U=2.25$ eV panel (b), the PM solution already features a definite Mott state.
(c,d) Momentum resolved spectral function for T = 193 K at $U=1.65$ eV (c) and $U=2.55$ eV(d). Experimental ARPES data are extracted from  Ref. \cite{Uchida_2014}.
    }
    \vspace{5pt}
    \hrule
    \label{fig:Appendix_5}
\end{figure}

\begin{figure}[t!]
    \centering
    \includegraphics[width=\linewidth]{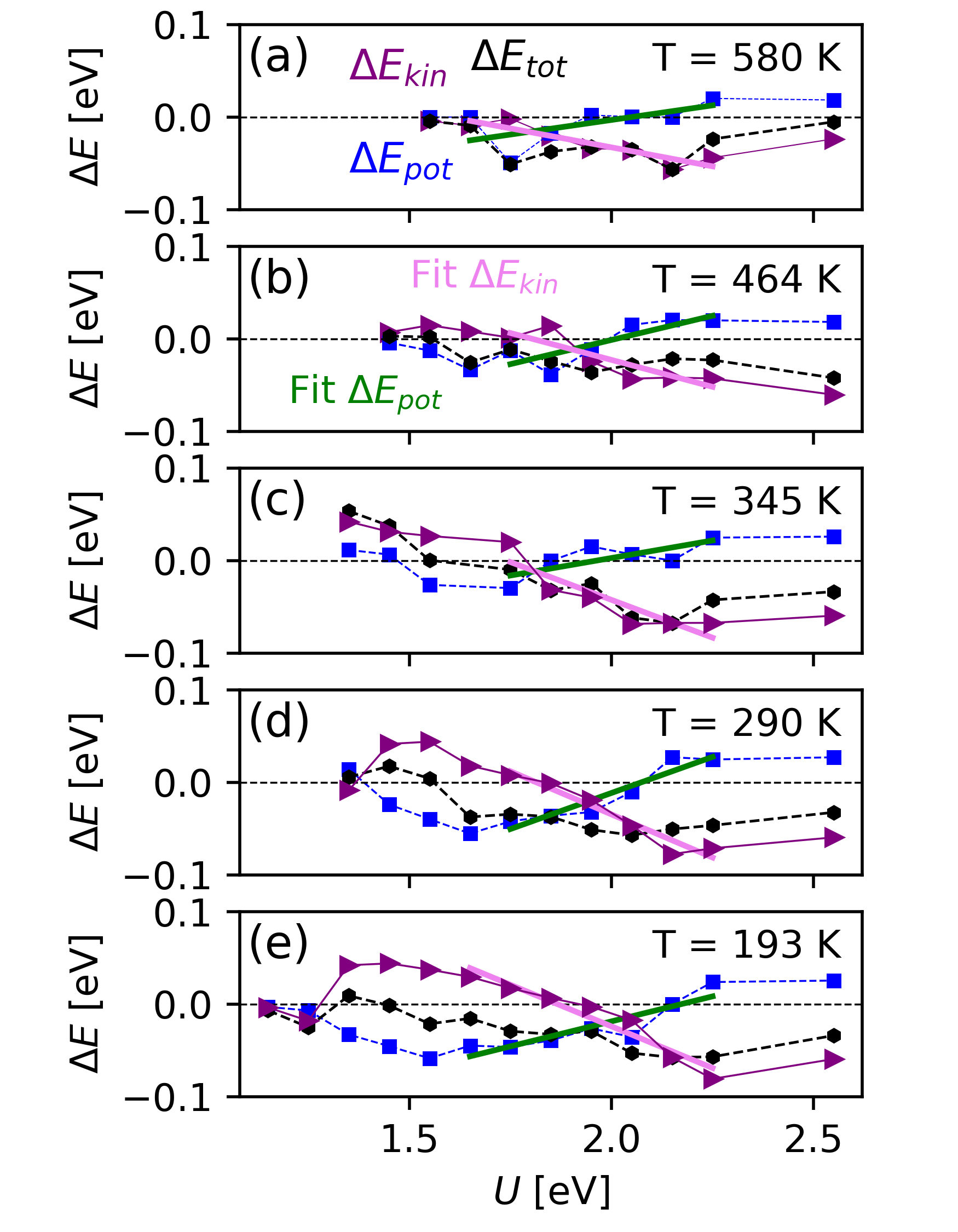}
    \caption{
Additional energy balance. Temperature evolution of $\Delta E_{kin}$, $\Delta E_{pot}$ and $ \Delta E_{tot}$. Green and light violet lines represent the linear fit of the kinetic and potential energy, respectively.  
    }
    \vspace{5pt}
    \hrule
    \label{fig:Appendix_6}
\end{figure}

\begin{figure}[t!]
    \centering
    \includegraphics[width=\linewidth]{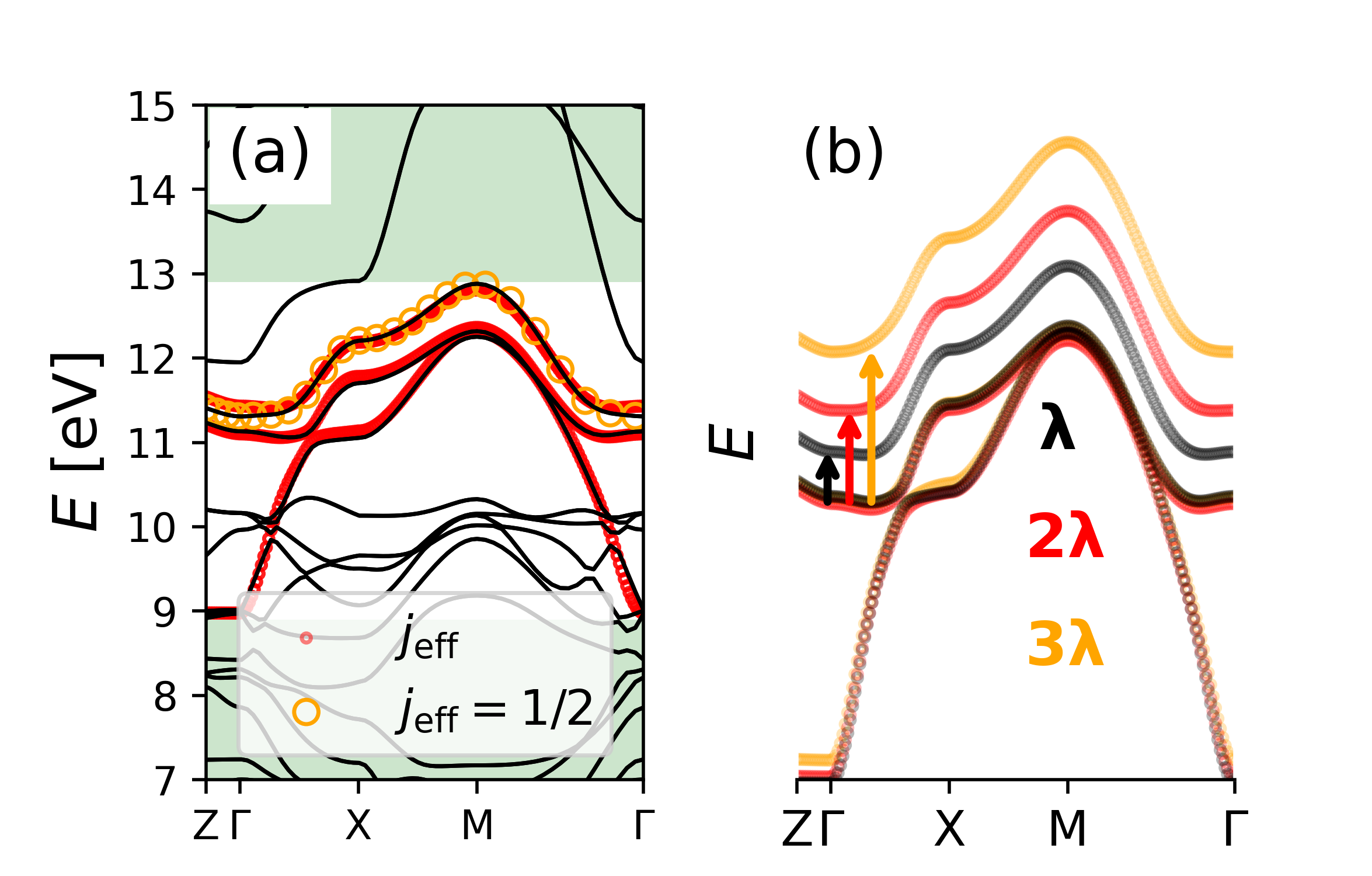}
    \caption{
DFT band structure and SOC. (a) DFT+SOC electronic structure and TB model for a three-orbital model with $\lambda = 0.31$ eV (red line) and TB obtained for the \jeff\ state only (orange curve). (b) Dependence of the band structure on $\lambda$ magnitude within the three-band model.
    }
    \vspace{5pt}
    \hrule
    \label{fig:Appendix_7}
\end{figure}

\newpage

\begin{table*}
\centering
{\renewcommand{\arraystretch}{1.5}
\begin{tabular}{c||l||c|c||c|c}
\toprule
& \multirow{2}{*}{$\mathbf{U}$ \textbf{[eV}]}  & \multicolumn{2}{c|}{\bairo}  & \multicolumn{2}{c}{\sriro} \\
\cmidrule(lr){3-4}\cmidrule(lr){5-6}
& & $U_{\mathrm{avg}}$ & $U_{1/2}$ & $U_{\mathrm{avg}}$ & $U_{1/2}$ \\
\midrule
\multirow{3}{*}{\Large\textbf{(i) Comparison to experiments}}
& ARPES & 1.62 - 2.52 & 1.65 - 2.55 & 1.65 - 2.55 & 1.61 - 2.52\\
& $\sigma(\omega) / \alpha(\omega)$ & 1.93 & 1.95 & 1.65 & 1.67  \\
& STS & / & / & 2.52 - 2.72 & 2.55 - 2.75 \\
\midrule
\multirow{5}{*}{\Large\textbf{(ii) \textit{Ab initio} estimates}} 
& Martins et \textit{al.}, 2011$^*$ & / & / & 2.25$^*$  & 2.20$^*$  \\
& Arita et \textit{al.}, 2012$^*$  & 1.63$^*$    &  1.65$^*$  & 1.96 & 1.93 \\
& Liu et \textit{al.}, 2018 & / & / & 1.82 & / \\
& Cassol et \textit{al.}, 2024 & 2.02 & 2.05 & / & / \\
& Zhang et \textit{al.}, 2025 & / & / & 1.97 & / \\
\midrule
\multirow{7}{*}{\makecell{\Large\textbf{(iii) Previous Theoretical} \\ \Large\textbf{works}}}
& Martins et \textit{al.}, 2011 &   / &  /   & 2.25  & 2.20 \\
& Arita et \textit{al.}, 2012   &  /    &  /   & 1.96 & 1.93 \\
& Liu et \textit{al.}, 2018 & / & / & 1.82 & / \\
& Zhang et \textit{al.}, 2021   &   /  &  /   & 2.4, 2.65 &  /   \\
& Zhang et \textit{al.}, 2023   &   /  &  /   & 2.65 &  /   \\
& Cassol et \textit{al.}, 2024  & 2.22 & 2.25 &  /   &  /   \\
& Zhang et \textit{al.}, 2025   &   /  &  /   & 2.65 &  /   \\
\bottomrule
\end{tabular}}
\caption{Supporting information of Figure \ref{fig:Figure_6}. The table summarizes: (i) Appropriate $U$ values to reproduce given experimental observables such as ARPES \cite{Moser_2014,Uchida_2014}, optical transport  \cite{Nichols_2014,Seo_2017}, and STS~\cite{Dai_2014}; (ii) \textit{ab initio} cRPA evaluations of the interaction terms for \bairo ~\cite{Arita_2012,Cassol_2024} and \sriro~\cite{Martins_2011,Arita_2012,Liu_2018,Pavarini_2025}; and (iii) Parameters employed in previous theoretical works to describe the physics of the compounds ~\cite{Martins_2011,Arita_2012,Liu_2018,Pavarini_2021,Pavarini_2023,Cassol_2024,Pavarini_2025}. In the cRPA sections, values marked with $*$ are calculated for a crystal structure that differs from the experimental one. Therefore, they are not reported in Figure ~\ref{fig:Figure_6}.
}\label{tab:Tab_1}
\end{table*}

\begin{table*}
    \centering
    {\renewcommand{\arraystretch}{1.5}
        \begin{tabular}{l||l||ccc||l||ccc}
             \toprule 
            & $U_{mm'}^{\sigma \bar{\sigma}}$ [eV] & $d_{xy}$ & $d_{yz}$ & $d_{xz}$ &  $U_{mm'}^{\sigma \sigma}$ [eV] & $d_{xy}$ & $d_{yz}$ & $d_{xz}$ \\
           \midrule       
           \multirow{3}{*}{\makecell{\Large  \textbf{Atomic} \\ \Large $\mathbf{t_{2g}}$ \textbf{basis}}}
             & $d_{xy}$ & 2.49 & 1.81 & 1.81 & $d_{xy}$ &  0.00 & 1.59 & 1.59 \\
             & $d_{yz}$ & 1.81 & 2.39 & 1.86 & $d_{yz}$ & 1.59 & 0.00 & 1.64 \\
             & $d_{xz}$ & 1.81 & 1.86 & 2.39 &  $d_{xz}$ &  1.59 & 1.64 & 0.00 \\
           \midrule    
            &  $U_{mm'}^{\sigma \bar{\sigma}}$ [eV] & $\ket{3/2,1/2}$  & $\ket{3/2,3/2}$ & $\ket{1/2,1/2}$ & $U_{mm'}^{\sigma \sigma}$ [eV] & $\ket{3/2,1/2}$  & $\ket{3/2,3/2}$ & $\ket{1/2,1/2}$ \\
            \midrule
            \multirow{3}{*}{\makecell{\Large  \textbf{Pseudospin} \\ \Large $\mathbf{j_{eff}}$ \textbf{basis}}}
            & $\ket{3/2,1/2}$ & 2.29 & 1.78 &1.72 & $\ket{3/2,1/2}$ & 0.00 & 1.67 & 1.81 \\
            &  $\ket{3/2,3/2}$ & 1.78 & 2.12 & 1.67 & $\ket{3/2,3/2}$ &1.67 & 0.00 & 2.04 \\
            & $\ket{1/2,1/2}$ & 1.72 & 1.67  & 2.05 & $\ket{1/2,1/2}$ & 1.81 & 2.04 & 0.00 \\
          \bottomrule 
        \end{tabular}}
        \caption{Local Coulomb interaction of \bairo\ within cRPA. In the atomic $t_{2g}$ basis, cRPA gives a generalized orbital dependent Hubbard Kanamori interaction, containing density-density, spin-flipping and pair-hopping terms. The repulsion between electrons with parallel spins is obtained as $U_{mm'}^{\sigma \sigma} = U_{mm'}^{\sigma \bar{\sigma}}- J$, where $J$ is the Hund's coupling $J = 0.22$ eV, which is orbital independent. In the $j_{\mathrm{eff}}$ basis, this relation is no longer satisfied, and the Coulomb tensor is composed of additional correlated hopping and four-scattering terms \cite{Kim_2020}. Notice that in this case $\sigma$ is not the canonical spin, but represents the projection of the total angular momentum $j$. Furthermore, the larger Coulomb interaction in the \JEFF\ subsector is due to symmetry properties of SOC  and not to the specific form of the Coulomb tensor in the $t_{2g}$ basis \cite{Martins_2017}.}
    \label{tab:Tab_2}
\end{table*}

\end{document}